\documentclass{aa}

\usepackage{graphicx}

\usepackage{txfonts}

\usepackage{natbib,twoopt}
\usepackage[colorlinks=true, linkcolor=blue, citecolor=blue, urlcolor=blue, breaklinks=true]{hyperref} 
\usepackage{orcidlink}

\usepackage{xcolor}

\begin{document} 

    \title{A search and multi-frequency study of curved parsec-scale jets in active galaxies}
    
    \author{
    V.~A.~Makeev\inst{1}\thanks{e-mail: makeev.va.ph@gmail.com; Member of the International Max Planck Research School for Astronomy and Astrophysics at the Universities of Bonn and Cologne}\orcidlink{0009-0008-7830-4553}
    \and A.~B.~Pushkarev\inst{2,3}\orcidlink{0000-0002-9702-2307}
    \and Y.~Y.~Kovalev\inst{1}\orcidlink{0000-0001-9303-3263}
    }
    
    \institute{
    Max-Planck-Institut für Radioastronomie, Auf dem Hügel 69, D-53121 Bonn, Germany
    \and
    Crimean Astrophysical Observatory, Nauchny 298688, Crimea, Russia
    \and
    Moscow Institute of Physics and Technology, Institutsky per. 9, Dolgoprudny 141700, Russia
    }
    
    \date{Received XXX / Accepted YYY}

    \abstract
    {Very long baseline interferometry (VLBI) observations show that a fraction of active galactic nucleus (AGN) jets exhibit pronounced apparent bending on parsec scales, potentially strongly enhanced by projection effects. While individual sources have been studied in detail, the statistical properties of curved jets and the physical mechanisms responsible for their morphology remain poorly constrained.}
    {We aim to select a statistically robust sample of AGNs with significantly curved parsec-scale jets and to constrain the physical mechanisms responsible for jet bending.}
    {We analyse 123\,456 VLBI images of 17\,478 AGNs collected by the public Astrogeo database. After filtering for extended structure, we extract and stack multi-epoch, multi-frequency (1.2--86~GHz) ridge lines. We fit linear and logarithmic helices representing jet precession and resonant Kelvin--Helmholtz instability mode. We combine model-fits and Monte Carlo simulations to further constrain these physical models.}
    {We robustly identify 586 AGNs (7.4\% of sources with resolved jets) with curved parsec-scale jets. Compared to straight jets, curved ones are less common at low redshifts, possibly due to the sample-related selection effects. In the precession scenario, the inferred periods span $P \sim 10\div10\,000$~yr, with a logarithmic mean of $\sim 220$~yr. Solid-body disc precession in supermassive black hole binaries is broadly consistent with these periods, whereas geodetic precession requires extremely tight, near-equal-mass systems. Lense--Thirring precession remains viable for both thin and thick discs. Kelvin--Helmholtz instability also provides a potential explanation. However, identifying the preferred model for each individual source requires further investigation, including temporal analysis and jet kinematics.
    }
    {}
    
   \maketitle
\nolinenumbers

    \section{Introduction}

        Decades of very long baseline interferometry (VLBI) observations of active galactic nuclei (AGNs) have shown that a significant fraction of AGN jets exhibit bending, magnified by projection effects. A study by \citet{Lister2021} on a sample of 419 jets shows that about 42\% of the 926 well-sampled features follow non-radial, curved trajectories, indicating the bending of the jets themselves since these features are expected to trace the outflow. Additionally, the study observed notable variations in the inner jet position angle, which may lead to pronounced bending on larger scales. This topic was further analysed by \citet{Kostrichkin2025}, who studied a sample of 317 AGN jets with notable variations in the direction of the VLBI jet. The variability was also connected with possible further jet bending. The next important signature of the jet bending is a misalignment of directions on parsec and kiloparsec scales. This phenomenon is demonstrated and studied by \citet{Pearson&Readhead1998} and \citet{Conway1993}, suggesting a helical structure of the jets. In fact, curved jets are often presumed to be helical and, as proposed by \citet{Steffen1997}, are categorised into three groups: ballistic helical jets, helically bent jets, and internally helical jets. Ballistic helical jets, such as B0605$-$085 \citep{Kudryavtseva2011}, involve individual elements like plasma blobs following straight paths but collectively forming a helical structure due to periodic changes in ejection direction. Helically bent jets, exemplified by PKS~2136$+$141 \citep{sbj3}, are bent as a whole with matter flowing along the helix. Internally helical jets may have internal matter flowing along several helical paths. Theoretical explanations for these phenomena include jet precession \citep{Roos1988, Fendt2022, Dubey2024} and distortions caused by plasma instabilities such as Kelvin-Helmholtz (KH), current-driven instabilities \citep{3dinstab} and kink instabilities \citep{Lalakos2024}. Besides helical structures, jets can follow irregularly bent paths, like local bends due to the interaction with surrounding medium \citep{ismcol, Sutherland2007, Choi2007}, often exaggerated by projection effects which may play a crucial role in our perception of the source. As proven by \citet{Kovalev2025} on the example of jet in the blazar PKS~1424$+$240, some sources may have critically small angles between the directions of their jets and our line of sight, which virtually makes us look inside the jet cone. In such cases we may expect projection effects to be maximised, resulting in more prominent apparent bending of structures inside the jet.

        As already mentioned, KH instability, driven by velocity shear between the jet flow and the surrounding medium, is a common cause of bent jet structures \citep{Hardee1987}. Examples of jets exhibiting such instabilities include 3C~273 \citep{Lobanov2001} and M87 \citep{Nikonov2023}. The KH instability may also originate from slow precession of the jet nozzle \citep{Lister2003}.
        
        The jet base precession may stem from two primary mechanisms: orbital revolutions within a supermassive black hole binary (SMBHB) system \citep{precbh} or the Lense-Thirring effect due to the misalignment of momenta vectors between a supermassive black hole (SMBH) and its accretion disc \citep{lentir}. The SMBHB hypothesis is particularly relevant due to the abundance of post-merger galaxies predicted by the hierarchical merger models of galaxy formation \citep{White1978}. After the merger, the central SMBHs move closer through dynamical friction and form close SMBHBs \citep{Mayer2007}. The model of SMBHB explains the variability and curvature in AGN jets like OJ~287 \citep{Liu2002}, 3C~345 \citep{Lobanov2005} and 3C~66A \citep{Thevenet2026}. Until the launch of the Laser Interferometer Space Antenna (LISA) \citep{lisa, Klein2016}, direct detection of binary black holes remains elusive, highlighting the importance of indirect methods such as studying AGN jet shapes. Alternatively, the disc-driven precession model, in which the black-hole spin is misaligned with the angular-momentum vector of the accretion disc, is supported by observations \citep{Caproni2004, Caproni2006} and general relativistic magnetohydrodynamic simulations \citep{Liska2017}. These jet structures offer insights into the dynamics and evolution of the central objects over significant time scales.

        In this paper, we present a search for curved AGN jets by studying a large sample of their VLBI AGN images. We analyse the geometrical properties of curved jets and elaborate on possible physical mechanisms that may cause bending. We put constraints on physical parameters of the SMBHB model and the accretion-disc-driven precession model. Additionally, we consider KH instability as a possible cause of bending.

        The paper is organised as follows: In Sect.~\ref{sec:obs_data} we describe VLBI data used in this study and initial filtering protocols. We present a search of significantly curved jets in Sect.~\ref{sec:search} and modelling of their geometrical shapes in Sect.~\ref{sec:geomodel}. We present the modelling results, physical interpretation, and discussion of the models in Sect.~\ref{sec:discussion} and summarise our conclusions in Sect.~\ref{sec:conclusion}. Throughout this paper we use the cosmological parameters $\Omega_m = 0.31$, $\Omega_\Lambda = 0.69$, and $H_0 = 67.4 \text{ km s}^{-1} \text{ Mpc}^{-1}$ \citep{Planck2020} and define sky PAs in degrees east of north.
        
    \section{Observational data} \label{sec:obs_data}
        \subsection{Data description}
            In this study, we use version 2020B of the Astrogeo VLBI FITS image database\footnote{Digital object identifier (DOI) \href{https://doi.org/10.25966/kyy8-yp57}{10.25966/kyy8-yp57}}, compiled from geodetic observations using VLBI, Very Long Baseline Array (VLBA) calibrator surveys  and a multitude of other VLBI observations \citep{Petrov2025}. It contains over 123~456 full-intensity images of 17~478 compact radio sources at frequencies ranging from 1.2 GHz to 86 GHz. On average, each object was observed in eight epochs and at two frequencies. Most images in the database do not have a resolved jet structure, partially due to the snapshot mode of a significant part of the observations, which renders the observations insufficiently sensitive to reveal the extended jet. To refine our analysis, we need to perform a filtering procedure and exclude unresolved sources at a given level of sensitivity from our sample.

        \subsection{Extended structure identification}

            To exclude images where the jet structure is not resolved or is not detected at a given level of sensitivity, we used the following algorithm:
            
            \begin{enumerate}
                \item The brightness distribution was modelled using a two-dimensional Gaussian function. 

                \item The model was subtracted from the brightness distribution.
                \item The residual flux density was calculated.
                \item If the remaining flux density exceeds the manually set optimal value ($\sigma_{\text{crit}}$), the target is considered to have an extended structure. We define this value in a way that ensures that among 200 randomly chosen images manually marked up, at least 90\% of the marked up images with an extended jet structure are automatically selected. 
            \end{enumerate}
            
            Indeed, due to the specifics of the CLEAN imaging algorithm, the restored images without an extended structure should correspond to a Gaussian distribution with an accuracy up to the noise level. To determine the parameter $\sigma_{\text{crit}}$, manual labelling of a sample consisting of about a hundred objects was conducted. The optimal value of $\sigma_{\text{crit}}$, at which we do not exclude too many resolved images while retaining only a few unresolved images in the sample, was determined to be
            $\sigma_{\text{crit}}=5.5 \sigma_{\text{map}}.$ The map noise, $\sigma_{\text{map}}$, was calculated as the median of the standard deviations measured in four square regions located in the image quadrants, each with a side length equal to 15\% of the image side. The filtering procedure reduced the initial data set by 54.6\%. The remaining part contains 63~879 images of 7~927 sources with a resolved jet structure.

    \section{A search for curved jets}\label{sec:search}
    
        Next, after applying the filtering procedure to the sample, it was necessary to identify objects with significantly distorted jets. To accomplish this task, we describe the spatial structure of the jet. One of the simplest ways is to associate each jet image with its ridge line. Qualitatively, the ridge line represents a curve that passes through the medians or modes of intensity distributions in cross-sections of the jet. In this study, we investigate the spatial variation of the geometry of the entire visible jet, rather than its internal structure, and therefore use medians.

        To construct the ridge lines, we used an algorithm similar to that described by \citet{Pushkarev2017}. The algorithm works as follows. For the first iteration, noise is subtracted from the map, numerically estimated as \(2 \sigma_{\text{map}}\). The brightest pixel in the map, typically located near its centre, is selected as the initial point of the ridge line. Concentric circles centred on this point are then constructed, and the intensity distributions of the pixels lying along these circles are analysed. Starting with the innermost circle, the median position of each intensity distribution is determined as follows. The circle is divided into two semicircles by a straight line oriented at an angle \(\varphi\), and the total intensity within each semicircle is calculated. The angle \(\varphi\) is then varied to find the value \(\varphi_0\) that minimises the relative difference between the total intensities of the two semicircles. The median position of the intensity distribution is defined by \(\varphi_0\).

        To estimate the uncertainties in each ridge-line point, we first assume the brightness distribution on the corresponding circle to be Gaussian of the form $I \approx A \exp{-(\phi^2 / 2\sigma^2)} + \sigma_{\text{map}} \cdot(r/\text{1 px})$, where $\rm 1\;px$ corresponds to the pixel decrement. Then, the intensity $F$ along the circle is computed. Under our assumption
        \begin{equation}
            F = \int I d\phi = \sqrt{2\pi}A\sigma + \sigma_{\text{map}} \cdot \frac{2\pi\, r}{\text{1 px}}
        \end{equation} 
        and the uncertainty in $\phi_0$ may be estimated as 
        \begin{equation}
            \sigma_\phi = \frac{F}{\sqrt{2\pi}A} = \sigma + \frac{\sigma_{\text{map}}}{A}\cdot \frac{\sqrt{2\pi}\, r}{\text{1 px}}.
    \end{equation}
        Such an error definition takes into account the width of the brightness distribution and the signal-to-noise ratio, $\text{SNR} \sim  \sigma_{\text{map}} / A \cdot(\sqrt{2\pi} r/\text{1 px})$. In the image plane, the errors along coordinate axes are estimated in the following way:
        \begin{equation}
            \begin{cases}
                \sigma_x &= |y|\sigma_\phi + |x|\frac{\Delta r}{r},\\
                \sigma_y &= |x|\sigma_\phi + |y|\frac{\Delta r}{r},
            \end{cases}
    \end{equation}
    where $\Delta r$ is a ridge line's radial step, which we set to be $1$~pixel.

        To search for candidates that potentially have a significantly curved jet structure, we overlay the obtained ridge lines and describe them using the linear function \(\phi = K r + \phi_0\), where \(r\) and \(\phi\) are the distance from the core and the position angle, respectively. Objects having a total change in angle \(\varphi\) in a linear approximation, 
        \begin{equation} \label{eq:simple_fit}
            \Delta \phi = |K| (r_{\text{max}} - r_{\text{min}}) > 20^{\circ},
        \end{equation}
        are considered significantly curved. The value of \(20^{\circ}\) was determined based on a manually annotated sample of 200 objects. This approach is particularly sensitive to sources with stable and prominent bending at most frequencies. If the jet has wiggles \citep[e.g. 4C~$+21.35$;][]{Lister2018}, the average change in position angle will be negligible, and the source will not pass the selection criteria. Another possible case is a jet that shows curvature on smaller scales but has negligible curvature on larger scales \citep[e.g. 0735$+$178;][]{Gabuzda2021}. This might also result in the source not being selected.
        
        To avoid uncertainty of the jet direction in the core region, \(r_{\text{min}} > 0\) is defined as the minimum radius of the concentric circle covering at least one pixel of the residual map (after noise subtraction) with zero intensity. To find \(r_{\text{max}}\), we calculate the average intensities on 5\% square segments of the map centred on each point of the ridge line. The largest radial distance at which the average intensity is greater than \(2 \sigma_{\text{noise}}\) defines \(r_{\text{max}}\). The angle \(\Delta\phi\) can be used only for a rough approximation to the visible jet bend angle since the morphologies of curved jets often significantly differ from those described by a linear relationship.
        
\begin{figure*}
            \centering
            \begin{minipage}[t]{0.25\linewidth}
                \center{\includegraphics[width = 1\linewidth]{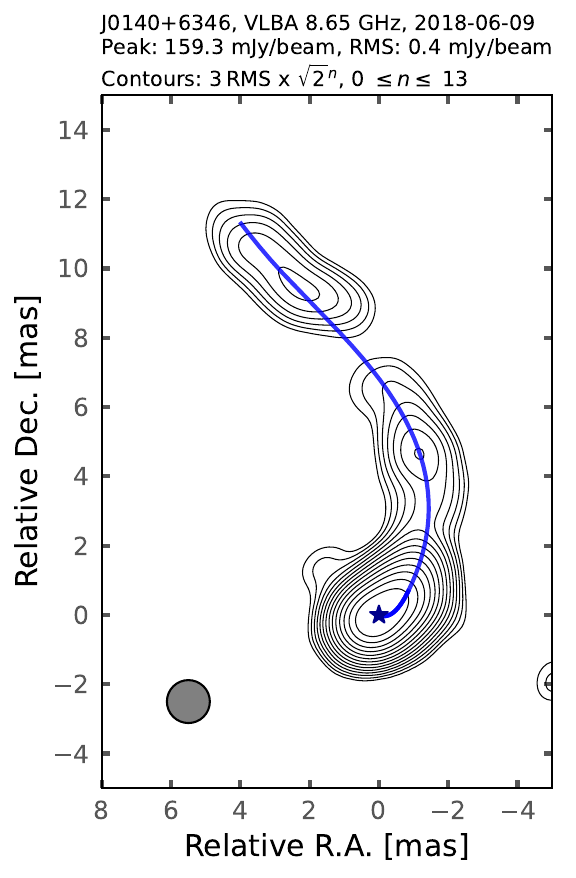}}
            \end{minipage}
            \hfill
            \begin{minipage}[t]{0.41\linewidth}
                \center{\includegraphics[width = 1.\linewidth]{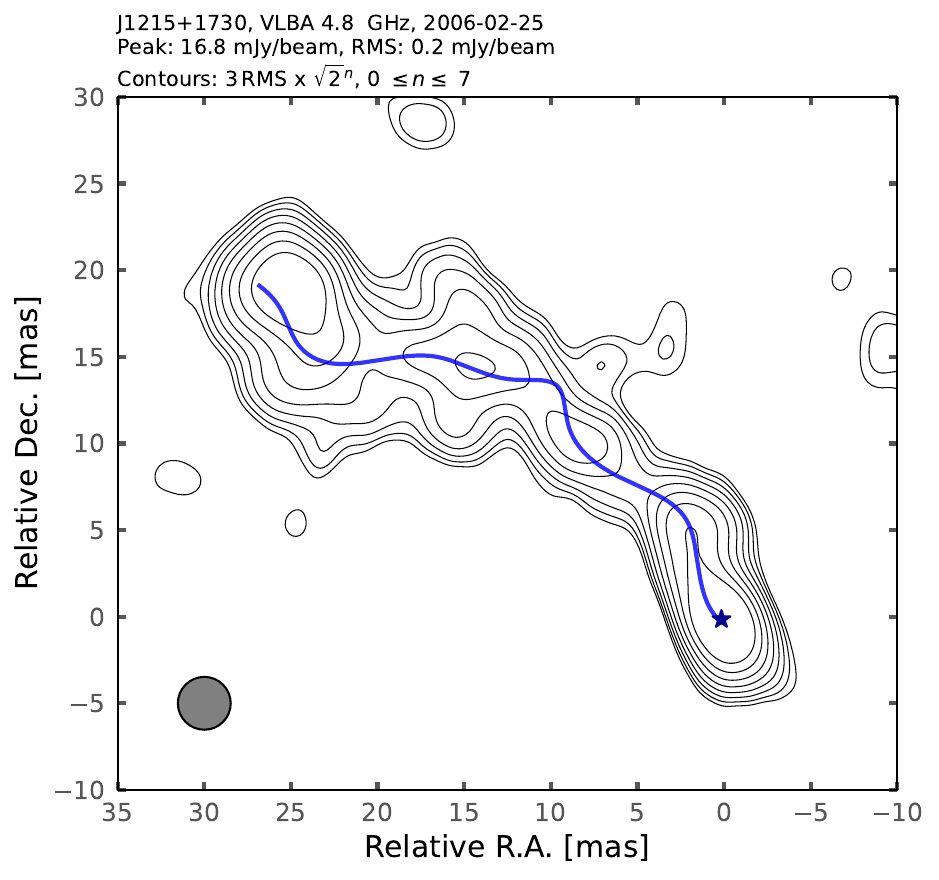}}
            \end{minipage}
            \hfill
            \begin{minipage}[t]{0.27\linewidth}
                \center{\includegraphics[width = 1\linewidth]{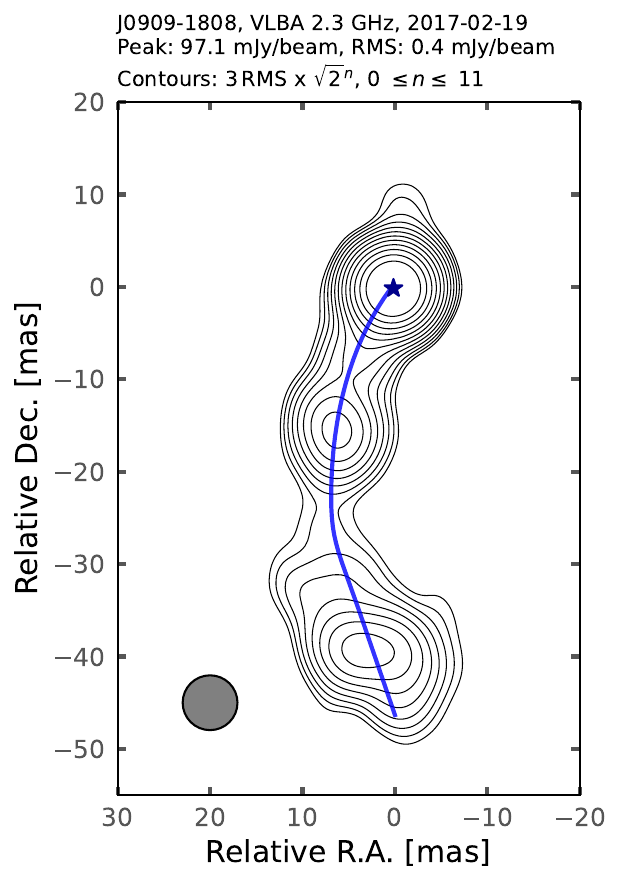}}
            \end{minipage}

            \caption{Examples of significantly curved jets. The contours are constructed at increasing powers of $\sqrt2$, starting at 3~RMS. The left panel shows J0140$+$6346, the middle panel shows J1215$+$1730, and the right panel shows J0909$-$1808. Peak intensities, frequencies, observation dates, and exact power decrements of the contours are indicated on the images. The brightest pixel locations are shown with an asterisk. The FWHM of the restoring beam for each image is shown in the bottom left corner as a shaded circle. Blue lines represent third-order splines fitted to the total-intensity ridge lines. }
            
            \label{fig:curved_example}
        \end{figure*}

        After the automatic selection of candidate sources with curved jets, a manual inspection was performed to remove objects that had been mistakenly selected because of poor map or ridge-line quality. In total, 1\,843 sources were reviewed during the final selection, of which 586, comprising 16\,273 individual maps, were classified as robustly curved (e.g. Fig. \ref{fig:curved_example}). The middle panel, showing the jet in J1215$+$1730, provides an example of ridge-line distortion near a relative right ascension of 10 mas. In this case, the ridge line is displaced from the centre of the jet channel by noise-induced emission on the right-hand side at the same radial distance. This is the most common type of error in the data. An additional 768 sources showed indications of jet curvature but were classified as non-robust because of the limited number or poor quality of the available maps. The names of both robust and non-robust sources are shown in Table~\ref{tab:curved_sources}.

        \begin{table}
            \centering
            \caption{General properties of the selected sample of AGN with curved jets.}
            \label{tab:curved_sources}
        
            \begin{tabular}{lccc}
                \hline
                \hline
                Source & $z$ & Opt. Class & Robust \\
                (1)    & (2) & (3)        & (4) \\
                \hline
                J0000$+$4054 & --    & -- & No  \\
                J0001$-$1551 & 2.044 & QSO & Yes \\
                J0002$+$2942 & --    & QSO & No  \\
                J0003$+$0717 & --    & -- & No  \\
                J0004$+$2637 & 0.500 & -- & No  \\
                J0005$+$3206 & --    & -- & No  \\
                \hline
                \multicolumn{4}{p{0.9\columnwidth}}{\footnotesize
                Columns are as follows: (1) Radio Fundamental Catalog (RFC) source name in J2000 notation; (2) redshift; (3) optical class; (4) curvature robustness flag (see Section~\ref{sec:search}). Redshifts are taken from OCARS as of August 2024 \citep{Malkin2018}, and optical classifications
                from \citet{Flesch2023}. The full table is available electronically from CDS; the first six entries are shown here for reference.}
            \end{tabular}
        \end{table}

\begin{figure*}
\centering
\includegraphics[height=0.32\linewidth]{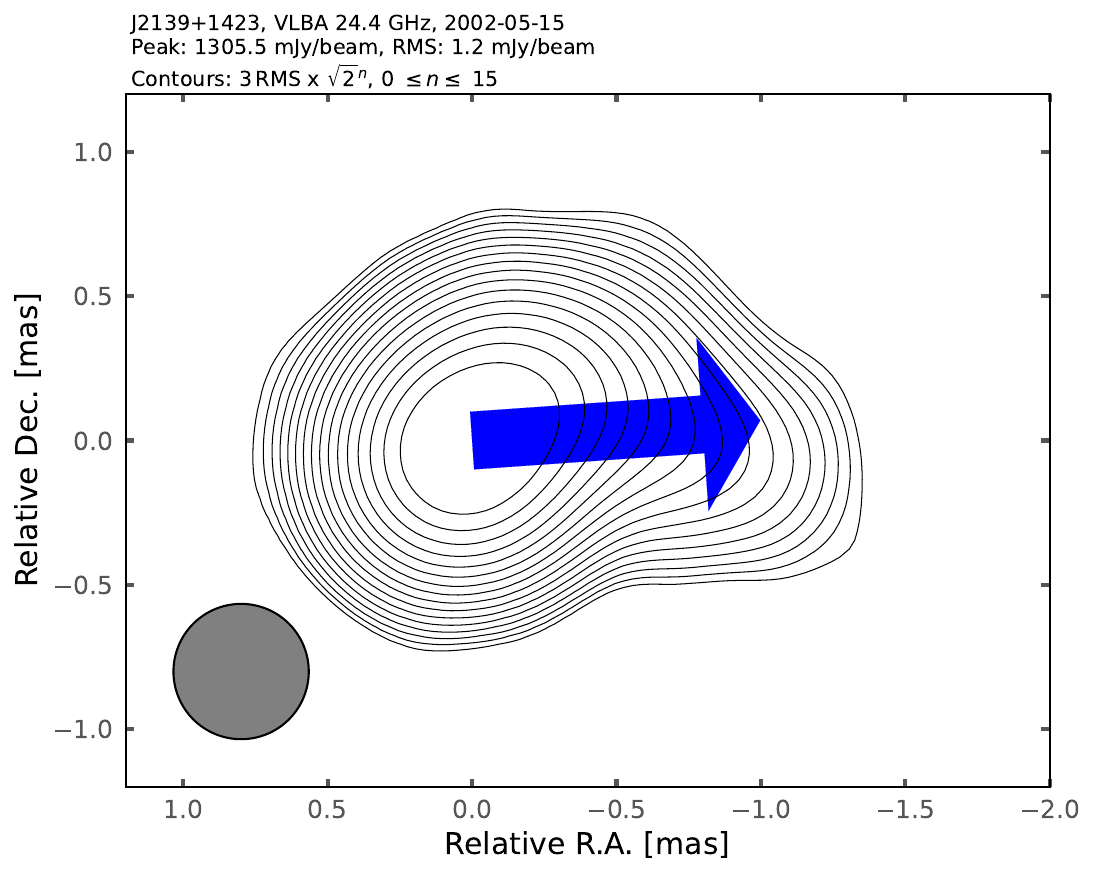}
\includegraphics[height=0.32\linewidth]{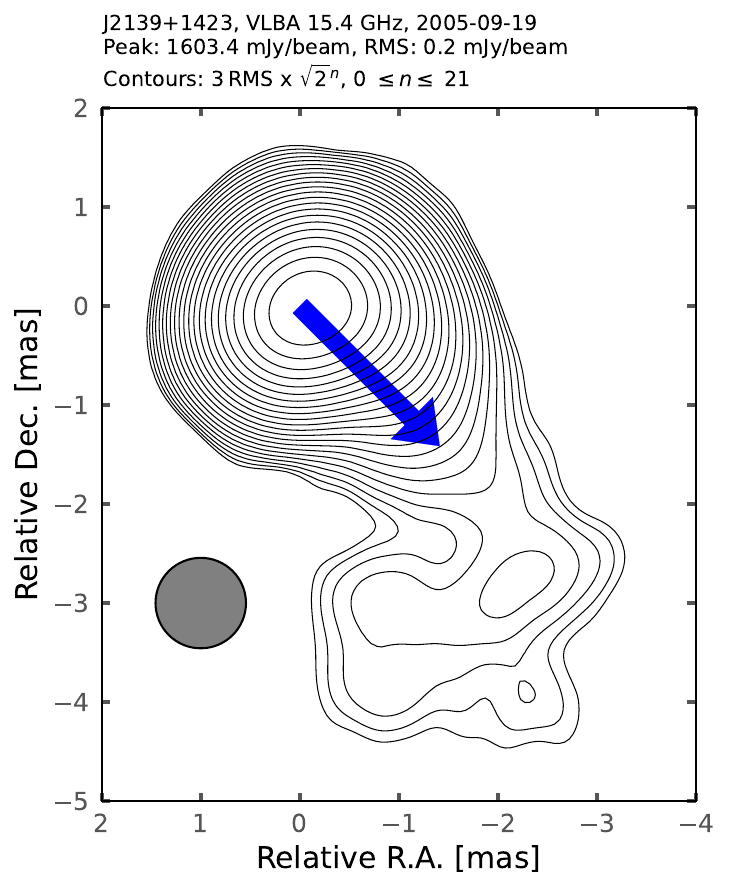}
\includegraphics[height=0.32\linewidth]{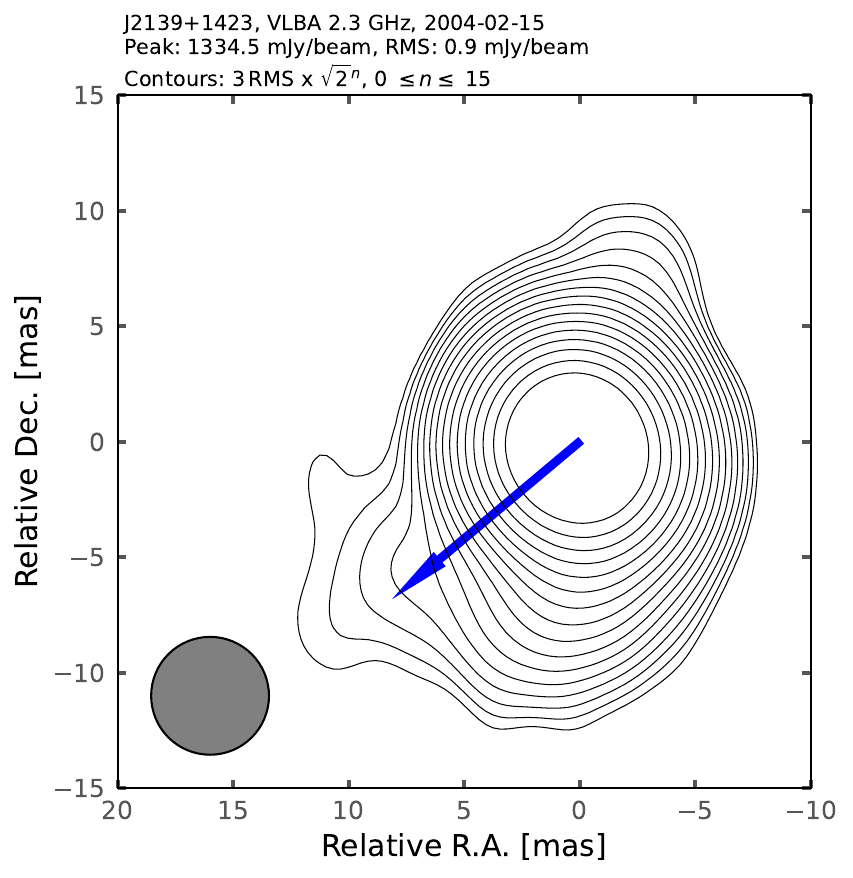}
\caption{The quasar J2139$+$1423 as an example of a source with jet direction changing with observing frequency, as probing different spatial scales. The contours are constructed at increasing powers of $\sqrt 2$.  Peak intensities, RMS, frequencies, observation dates and exact power increments of the contours are indicated on the images. The restoring diagrams for each image are shown in bottom left corner as shaded circles. Blue arrows represent visible directions of the jets at the given epochs.}
\label{fig:diff_dir_1}
\end{figure*}

\begin{figure*}
            \centering
            
            \begin{minipage}[t]{0.55\linewidth}
                \center{\includegraphics[width = 1\linewidth]{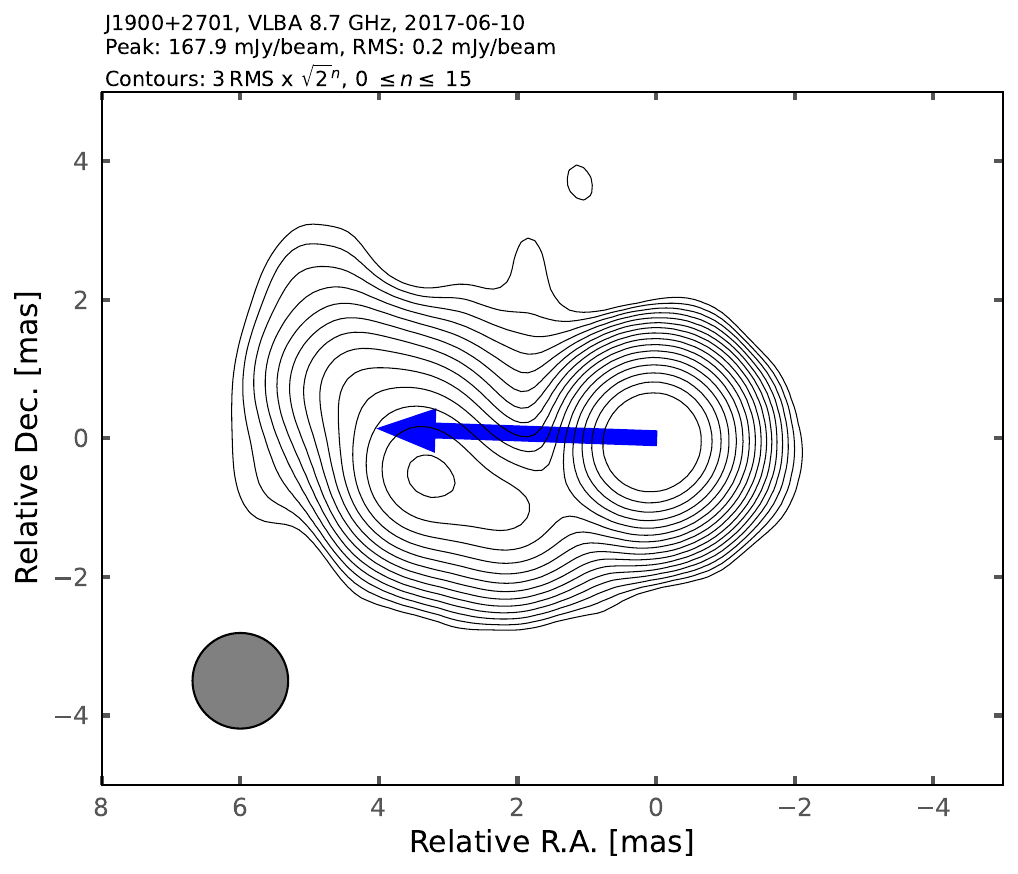}}
            \end{minipage}
            \hfill
            \begin{minipage}[t]{0.40\linewidth}
                \center{\includegraphics[width = 1\linewidth]{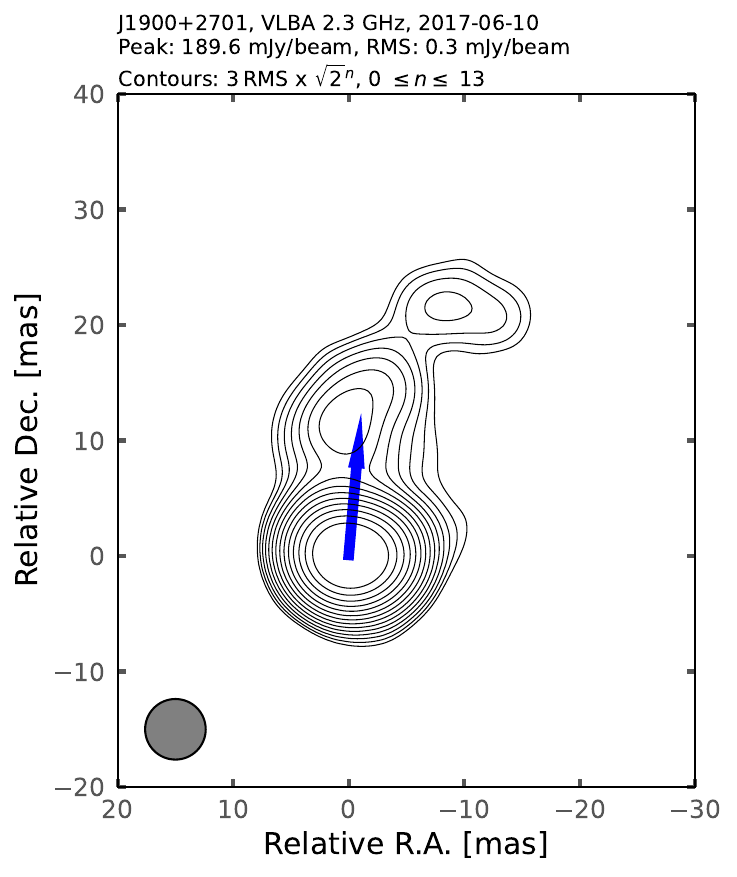}}
            \end{minipage}

            \caption{Examples of jet of the quasar J1900$+$2701 directed differently at different frequencies. The contours are constructed at increasing powers of $\sqrt2$ starting with the 3~RMS.  Peak intensities, RMS, frequencies, observation dates and exact power increments of contours are indicated on the images. The FWHM of the restoring beams for each image are shown in the bottom left corner as shaded circles. Blue arrows represent visible trend in directions of the jets at the given epochs.}
            
            \label{fig:diff_dir_2}
        \end{figure*}

        Median number of selected observing VLBI epochs per source is 11 (with mean 27) and median number of unique frequencies is 3 (with mean 3). X-band ($\sim$8~GHz) observations are present for 97\% of sources, S-band ($\sim$2.4~GHz) observations for 87\%, C-band ($\sim$5~GHz) for 52\%, U-band ($\sim$15~GHz) for 32\%, and L-band ($\sim$2~GHz), K-band ($\sim$22~GHz), Q-band ($\sim$43~GHz), W-band ($\sim$86~GHz) for less than 10\%. The importance of using multiple frequencies rather than one can be demonstrated by images of jets of quasars J2139$+$1423 and J1900$+$2701 shown in Fig. \ref{fig:diff_dir_1} and \ref{fig:diff_dir_2} accordingly. Both sources do not have visible significantly curved structure on separate frequencies. However, the position angle of the jet changes drastically at different frequencies as we observe larger structure with decreasing frequency due to the synchrotron self-absorption effect \citep{Blandford1979, Lobanov1998, Pushkarev2012}, as well as generally steep spectrum of optically thin jet emission \citep{Hovatta2014}. For J1900$+$2701 the visible part of the jet rotates by approximately 90$^\circ$ and for J2139$+$1423 the jet rotates by more than 120$^\circ$.  This change of position angle strongly indicates substantial curvature of the jet as a whole. It is not revealed by a restored map at a given frequency due to limited sensitivity and angular resolution of the array. 
        
        
\begin{figure}
            \centering
            \includegraphics[width=1.0\linewidth]{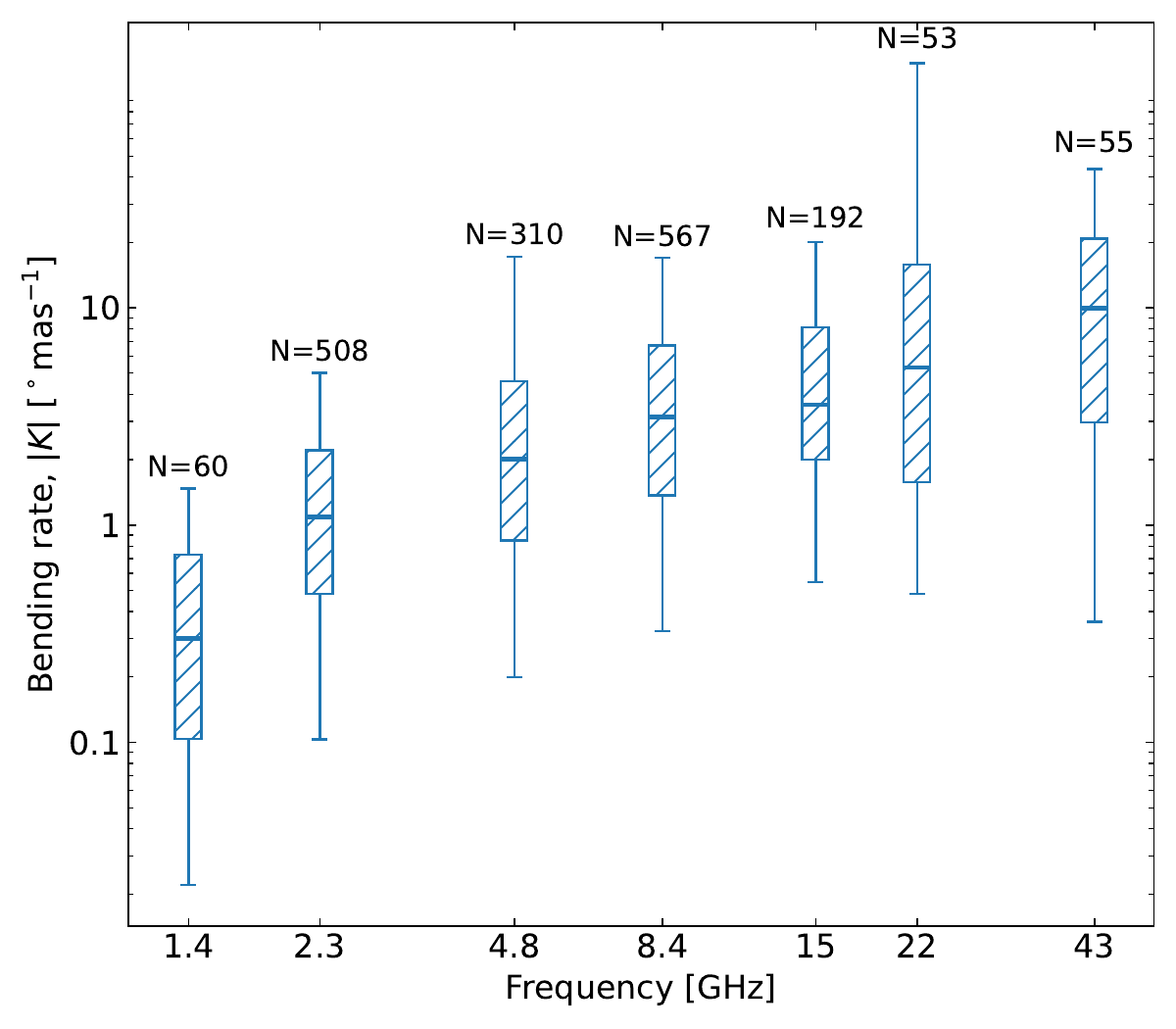}
            \caption{Frequency dependence of the jet bending rate for the significantly curved sources. The bending rate is quantified by the absolute value of the linear position-angle gradient, $|K|$, defined in Eq.~\ref{eq:simple_fit}. For each frequency band, the box shows the 25th--75th percentile range of $|K|$, the horizontal line marks the median, and the whiskers extend from the 5th to the 95th percentile. The number above each distribution gives the number of sources with available measurements in the corresponding band.}
            \label{fig:freq_bending}
        \end{figure}

        We additionally investigate the frequency dependence of jet bending. For each source, we stack all available ridgelines within the same frequency band and fit the resulting stacked ridgeline using Eq.~\ref{eq:simple_fit}. The derived bending rate shows a clear increase towards higher observing frequencies, indicating that the strongest changes in jet direction occur in the innermost regions probed at high frequencies (Fig.~\ref{fig:freq_bending}). This suggests that the overall change of the jet position angle with distance is generally not linear. The bending measurements obtained independently for each frequency band, together with the value derived from the combined multi-frequency ridgeline stack, are listed for individual sources in Table~\ref{tab:bending_freq}.

        \begin{table}
\caption{Frequency-dependent jet bending properties for the first six sources in the curved-jet sample.}
\label{tab:bending_freq}

\centering
\small

\begin{tabular*}{\columnwidth}
{@{\extracolsep{\fill}} l c c c c c @{}}

\hline\hline

Source &
Band &
$\nu$ &
$|K|$ &
$\Delta{\rm PA}_{\rm lin}$ &
$\Delta r$ \\

& &
[GHz] &
[deg\,mas$^{-1}$] &
[deg] &
[mas] \\

(1) & (2) & (3) & (4) & (5) & (6) \\

\hline

J0001$-$1551
    & S   & 2.3 & 1.090 & 31.6 & 28.99 \\
    & X   & 8.4 & 8.478 & 64.5 &  7.61 \\
    & All & --  & 1.743 & 50.9 & 29.19 \\\\

J0008+1144
    & S   & 2.3 & 0.424 & 19.3 & 45.43 \\
    & X   & 8.4 & 0.905 &  5.3 &  5.90 \\
    & All & --  & 0.611 & 28.0 & 45.77 \\\\

J0008+6837
    & S   & 2.3 & 4.078 & 92.8 & 22.75 \\
    & X   & 8.4 & 0.167 &  2.9 & 17.18 \\
    & All & --  & 2.239 & 50.9 & 22.75 \\\\

J0019+7327
    & L   & 1.4  &  0.099 &  7.9 & 80.45 \\
    & S   & 2.3  &  0.402 &  8.3 & 20.59 \\
    & C   & 4.8  &  0.713 & 18.0 & 25.26 \\
    & X   & 8.4  &  2.758 & 52.1 & 18.87 \\
    & U   & 15.0 & 12.755 & 61.1 &  4.79 \\
    & All & --   &  0.457 & 42.3 & 92.39 \\\\

J0023+0101
    & C   & 4.8 & 7.005 & 34.6 & 4.94 \\
    & All & --  & 7.005 & 34.6 & 4.94 \\\\

J0024+4644
    & S   & 2.3 & 0.399 & 13.7 & 34.24 \\
    & X   & 8.4 & 1.763 & 18.7 & 10.60 \\
    & All & --  & 0.534 & 21.0 & 39.35 \\

\hline

\multicolumn{6}{@{}p{\columnwidth}@{}}{
\footnotesize
Columns are as follows: (1) source name; (2) observing band;
(3) representative observing frequency; (4) absolute value of the
linear position-angle gradient $K$ defined in
Eq.~\ref{eq:simple_fit}; (5) total position-angle change from the
linear fit, $\Delta{\rm PA}_{\rm lin}=|K|\Delta r$; (6) radial span
$\Delta r=r_{\max}-r_{\min}$. Rows labelled ``All'' correspond to
fits to the combined multi-frequency ridgeline stack. The full table
is available electronically from CDS; the first six sources are shown
here for reference.
}

\end{tabular*}

\end{table}
        
        \section{Redshift and Optical Characteristics of the Sources}
        
\begin{figure}[ht]
            \centering
        
            \includegraphics[width=\linewidth]{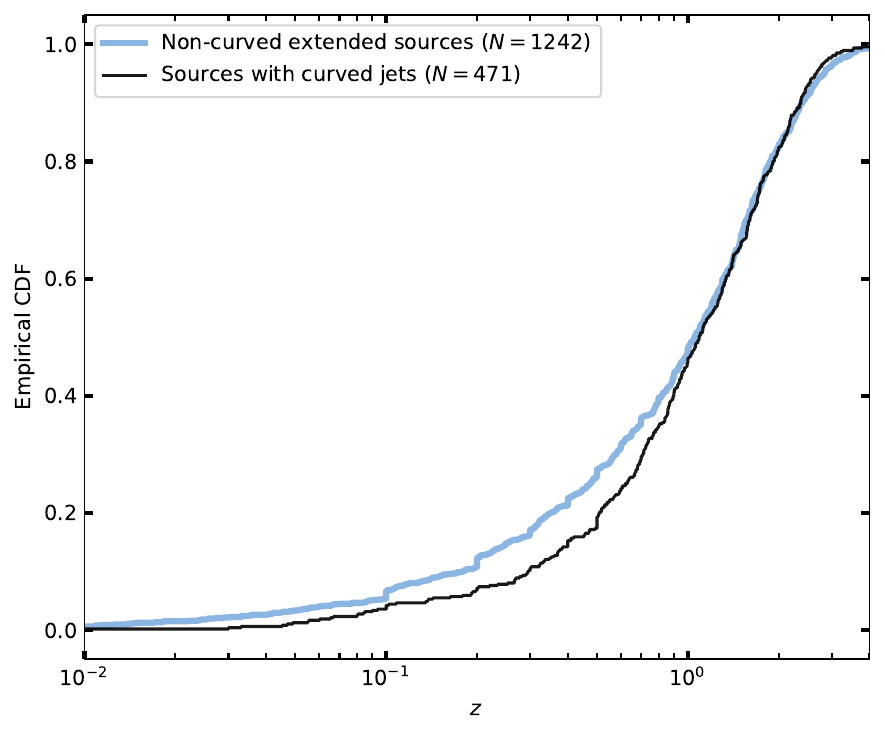}

            \caption{Cumulative distribution of redshift for sources with curved jets and sources with extended structure. Numbers of sources with known $z$ in each group, as well as the total numbers of sources, are shown in the plot legend.}
            \label{fig:z_dist}
        \end{figure}

        To study the redshift distribution of the sources in our shortlist, we utilise the OCARS catalogue\footnote{\href{https://www.gaoran.ru/english/as/ac\_vlbi/\#OCARS}{https://www.gaoran.ru/english/as/ac\_vlbi/\#OCARS}} \citep{Malkin2018}, which provides redshifts for 80.4\% of the sources with curved jets and 58.6\% of the remaining sources with extended jets. We show cumulative distribution functions (CDFs) of $z$ in Fig.~\ref{fig:z_dist}. There is a significant difference between the redshift distributions of sources with curved jets and all remaining extended sources. The Kolmogorov--Smirnov statistic is $0.088$, and the corresponding p-value is $p = 0.0089$. The CDF shows that sources with curved jets are less common at lower redshifts ($z < 1$) and slightly more prevalent at higher redshifts ($z > 1$) compared to sources without significant jet curvature. This observation is consistent with the following selection bias: assuming the same angular resolution and observing frequency, more distant sources are studied with poorer linear resolution. This means that we probe more extended jet structures, where apparent curvature may become more pronounced even if the intrinsic jet geometry is similar. Another possible explanation is that distant sources are more strongly dominated by beamed objects, implying smaller viewing angles and stronger projection effects, which can make jet curvature appear more prominent.  
        
\begin{figure}
            \centering
            \includegraphics[width=\linewidth]{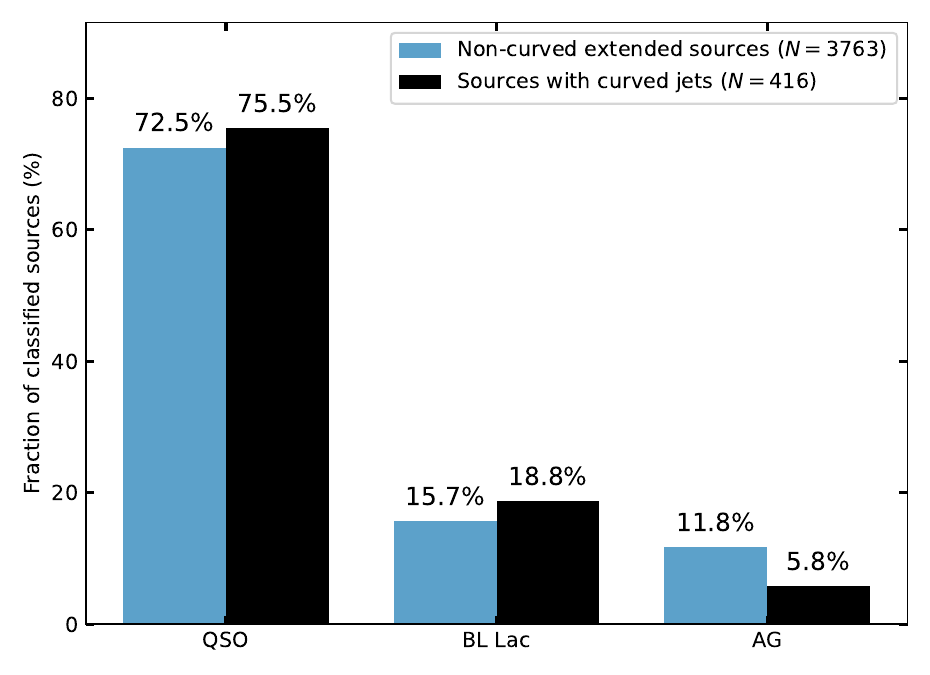}
            \caption{Optical classification of extended sources with curved and non-curved jets. The numbers of sources with known optical classifications in each group, as well as the total numbers of sources, are shown in the plot legend. The AG category includes host-dominated type-1 AGNs and type-2 narrow-line AGNs, while QSO represents quasi-stellar objects (type-1 broad-line, core-dominated sources).}
            \label{fig:optical}
        \end{figure}

        For optical classification, we use the Million Quasars catalogue \citep[Milliquas v8;][]{Flesch2023}, which provides a uniform classification scheme. We group its classes into three categories: active galaxies (AG; type-1 host-dominated AGNs, such as Seyfert~1 galaxies, and type-2 narrow-line AGNs, including type-2 QSOs), BL~Lacertae objects (BL~Lac), and quasi-stellar Objects (QSO; type-1 broad-line, core-dominated sources). The distribution of optical classes is shown in Fig.~\ref{fig:optical}. Of all extended sources, 3,763~(47\%) have optical classifications. The classification completeness is higher among sources with curved jets (71.0\%, 416 out of 586) than among sources with non-curved jets (45.6\%). Based on the two-sample z-test for proportions, there is a significant difference between the fractions of active galaxies ($s=-4.04$, $p=5.3\times10^{-5}$) and weak evidence for an excess of BL~Lacs ($s=1.83$, $p=0.068$), while no significant difference is found for QSOs ($s=1.43$, $p=0.15$). For sources with curved jets, the relative fraction of active galaxies is 54\% lower than for sources with non-curved jets (5.8\% versus 12.5\%), while the relative fraction of BL~Lacs is 23\% higher (18.8\% versus 15.3\%).

        The deficit of active galaxies can be largely explained by their generally larger jet viewing angles \citep{Urry1995,Homan2021}, since intrinsic bends appear less amplified by projection. Radio galaxies and other less-beamed AGNs are also typically found at lower redshifts, while nearby sources are underrepresented in our curved-jet sample; therefore, this selection effect may further reduce the fraction of active galaxies. The apparent excess of BL~Lacs might partly be related to the same projection effects, since their jets are viewed at smaller angles and modest intrinsic bends can appear strongly amplified \citep{Urry1995,Homan2021}. Strong pc-scale bending, jet wobbling, and large-scale misalignments have indeed been reported in BL~Lac jets \citep{Appl1996,Pushkarev2017,Britzen2001,Britzen2010}, and their strongly beamed, bright one-sided jets can also make curvature easier to detect \citep{Lister2013}. However, because the optical classifications are incomplete, with different completeness between the two samples, the apparent BL~Lac excess, as well as the absence of a corresponding difference for QSOs, may also be partly influenced by less obvious selection biases rather than solely reflecting intrinsic physical differences between the optical classes.

    \section{Modelling the geometry of jets}\label{sec:geomodel}

        \subsection{Morphological model: precession}
            The presented equations describe the ballistic motion of matter in a jet that undergoes precession. In simple terms, the base of the jet rotates, causing the matter stream ejected from it to follow a helix. Following \citet{Gower1982}, the velocity components of matter may be parametrised as:
            \begin{equation}
                \begin{cases}
                    \beta_x(t) = \beta \left[ \sin \psi \sin i \cos(\Omega(t_{\text{ej}} - t_{\text{ref}})) + \cos \psi \cos i \right], \\

                    \beta_y(t) = \beta \sin \psi \sin (\Omega(t_{\text{ej}} - t_{\text{ref}})),\\

                    \beta_z(t) = \beta \left[ \cos \psi \sin i - \sin \psi \cos i \cos (\Omega(t_{\text{ej}} - t_{\text{ref}})) \right],
                \end{cases}
            \end{equation}
            where $\Omega = 2\pi / P$ ($P$ is the precession period), $t_{\text{ej}}$~-- ejection time of the specific matter volume, $t_{\text{ref}}$~-- reference time, $\psi$ -- precession cone half-opening angle and $i$ is the angle between the precession cone's symmetry axis and the line of sight. Plasma speed and its components in light speed units are defined as $\beta$ and $\beta_x$, $\beta_y$, $\beta_z$.
            
            The corresponding relative angular coordinates are expressed as
            \begin{equation}
                \begin{cases}
                \displaystyle Z = \beta_z (t_1 - t_{\text{ej}}) \frac{c}{d}\frac{1}{1 - \beta_x},\\
                \displaystyle Y = \beta_y (t_1 - t_{\text{ej}}) \frac{c}{d}\frac{1}{1 - \beta_x},
                \end{cases}
            \end{equation}
            where $d$ is the distance to the source and $t$ is time in the source's frame.
            
            Assuming that the angle between the precession cone axis and the line of sight is small ($i \ll 1$) and that the velocity of the jet along the precession axis dominates over the other components ($\beta_z^2 + \beta_y^2 \ll \beta_x^2$), we obtain the following helix equations, parametrised by the angle $\phi$, such that $\Omega(t_{\text{ej}} - t_{\text{ref}}) = \Theta - \Omega(t_1 - t_{\text{ej}}) = \Theta - \phi$, with phase~$\Theta$:
            \begin{equation}
                \begin{cases}
                Z = K \sin (\Theta - \phi) \phi,\\
                Y = K [\alpha - \cos(\Theta - \phi)]\phi,
                \end{cases}
            \end{equation}
            where we use the following substitutions:
            \begin{equation} \label{eq:k_param}
                \begin{aligned}
                K &= \frac{1}{\pi} \frac{c}{d} \frac{\beta}{1 - \beta} P \sin \psi,\\
                \alpha &= \frac{i}{\tan \psi}
                \end{aligned}
            \end{equation}
            Additionally, we consider the position angle of the source's frame in the plane of the sky, which introduces an additional rotation of the $PA$. The final equations of the helix on the plane of the sky are:
                        \begin{equation}\label{eq:helix_lin}
                \begin{cases}
                    Z' = K \left[\cos(\Theta + \text{PA} - s\cdot\phi) - \alpha \sin \text{PA} \right] \phi,\\
                    Y' = K\left[-\sin(\Theta + \text{PA} - s\cdot\phi) + \alpha \cos \text{PA}  \right]\phi,
                \end{cases}
        \end{equation} 
        where we introduce an additional parameter $s = \pm1$ that defines the rotation direction of the jet base. 
        
        In order to properly model observed helices, two cases should be studied separately: (i) the viewing angle exceeds the precession cone angle ($\alpha\sim 1$), (ii) the line of sight is inside the precession cone ($\alpha \ll 1$). In the second case, we may opt for the helix parametrisation in terms of the position angle $\phi$ and distance from the core $r$: 
        \begin{equation}\label{eq:helix_lin_0alpha}
                        \phi = \phi_0 + K^{-1}r,
        \end{equation}
        where $\phi_0$ is the initial phase of the helix.

        \subsection{Morphological model: instabilities}\label{sec:morph_inst}
        
            As described in \citet{hardee06}, jets in AGN can be subject to current-driven instability and KH instability. Current instability occurs in magnetically dominated regions of the jet, while KH instability may develop in kinetically dominated regions of the jet. As the magnetic field in the jet decreases with distance from the central engine due to jet expansion, the kinetically dominated mode and, consequently, KH instability are expected on parsec scales. Plasma instabilities are widely studied using numerical simulations \citep{KH, 3dinstab}, but there is no model that analytically describes in three dimensions the complete structure of an unstable jet. However, in the linearised approximation, models can be obtained that still provide the opportunity to quantitatively describe the morphology of such a jet. As shown in \citet{KH}, the resonant wavelength of the KH instability, \(\lambda\), normalised to the jet radius \(r\), should remain constant:
            
            \begin{equation}
                \frac{\lambda}{r} \approx {\rm const}.
            \end{equation}       
            Furthermore, assuming a constant jet opening angle, \(\psi=\mathrm{const}\) \citep{hardee06}, and following the approach of \citet{KH, sbj3}, we assume that the wavelength of the dominant KH instability mode increases linearly with distance,
            $
            \lambda(z)=\lambda_0 (r/r_0)
            =\lambda_0+k(z-z_0).
            $
            This yields the following expression for the evolution of the mode with distance:

            \begin{equation}\label{eq:Hardee_int}
                 \varphi = \varphi_0 \pm 2\pi\int_{z_0}^{z} \frac{\text{d}z}{\lambda(z)},
            \end{equation}
            where \(\varphi_0\) is the initial phase. Using the assumptions of our model, we have:
            
            \begin{equation}\label{eq:KH_shape}
                \varphi = \varphi_0 \pm \frac{2 \pi \; r_0 }{ \tan{\psi \lambda_0}} \ln \frac{r}{r_0},
            \end{equation}
            or in terms of the sky plane coordinates, 
                        \begin{equation}
                \begin{cases}
                    Z' = R_0\left[cos(\Theta + \text{PA} - s\cdot\phi) - \alpha \sin \text{PA} \right] \exp\left(\frac{\phi}{K_\text{log}}\right),\\
                    Y' = R_0\left[-\sin(\Theta + \text{PA} - s\cdot\phi) + \alpha \cos \text{PA}  \right]\exp\left(\frac{\phi}{K_\text{log}}\right),
                \end{cases}
        \end{equation} 
        where we set $R_0 = 1$~mas, absorbing the corresponding scale change into the parametrisation of $\phi$, and 
        \begin{equation}
            K_\text{log} = \frac{2\pi}{\tan{\psi} }\left[\frac{r_0}{\lambda_0}\right].
        \end{equation}

        \subsection{Model parameter estimation}
        \label{subsec:MCmodelling}

            The linear- and logarithmic-helix models were fitted to the stacked
            ridge lines using global non-gradient optimisation. Details of the
            fitting procedure, parameter boundaries, and uncertainty estimation
            are provided in Appendix~\ref{sec:model_fitting_appendix}.

        The main parameter for both linear and logarithmic helices that contains information about the physics of the source is \( K \). For a linear helix, \( K \) is constructed from \( d \), \( \beta \), \( \psi \), and \( P \). If we know the values for the first three, we can estimate the period. Unfortunately, for our dataset of curved objects, we are not yet able to estimate these parameters directly. To overcome this problem, we use a Monte Carlo simulation approach. This approach suggests that if we know the prior distributions of the initial parameters, we can establish the posterior distribution of the secondary parameter related to them. Knowledge of \( K \) establishes this relationship (Eq. \ref{eq:k_param}).

Further, we assume that in our dataset, the parameters \( \beta \) and \( \psi \) follow the same distribution as those from the MOJAVE sample \citep{Pushkarev2017} and are statistically independent. Specifically, for the angle \( \psi \), we assume it has the same order of magnitude as the intrinsic jet's opening angles and thus follows the same distribution. To construct the posterior distribution of \( P \) for each source, we randomly sample the values of these parameters from the MOJAVE distributions and the observed $z$ distribution for curved jets from OCARS data. The parameter $d$ is computed from $z$ as the luminosity distance. For the sampling procedure we fit the observed distributions with the generalised lambda distribution (GLD). Fitting details are presented in \autoref{sec:gld}. We choose a total of 10~000 random parameter sets for each Monte Carlo model.

To establish a posterior distribution of \( \lambda_0/r_0 \) for the logarithmic helix model, we follow the same approach, with the only parameter modelled being the angle \( \psi \).

    \section{Results and Discussion} 
    \label{sec:discussion}
    
        \begin{figure*}
        \centering

        \begin{minipage}[t]{0.32\linewidth}
            \centering
            \includegraphics[width=\linewidth]{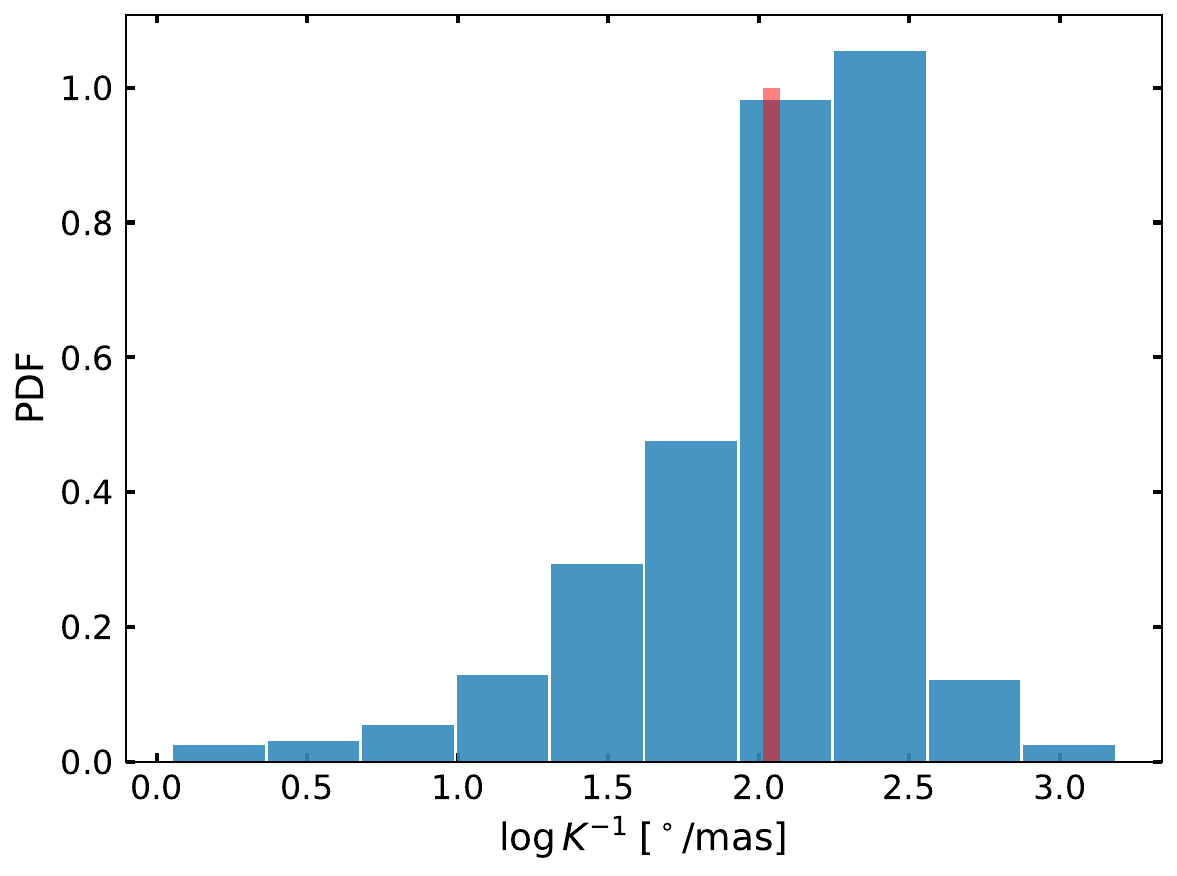}\\[0.3em]
            {\small (a) Linear helix model ($\alpha = 0$)}
            \label{fig:K_M0_lin_a}
        \end{minipage}
        \hfill
        \begin{minipage}[t]{0.32\linewidth}
            \centering
            \includegraphics[width=\linewidth]{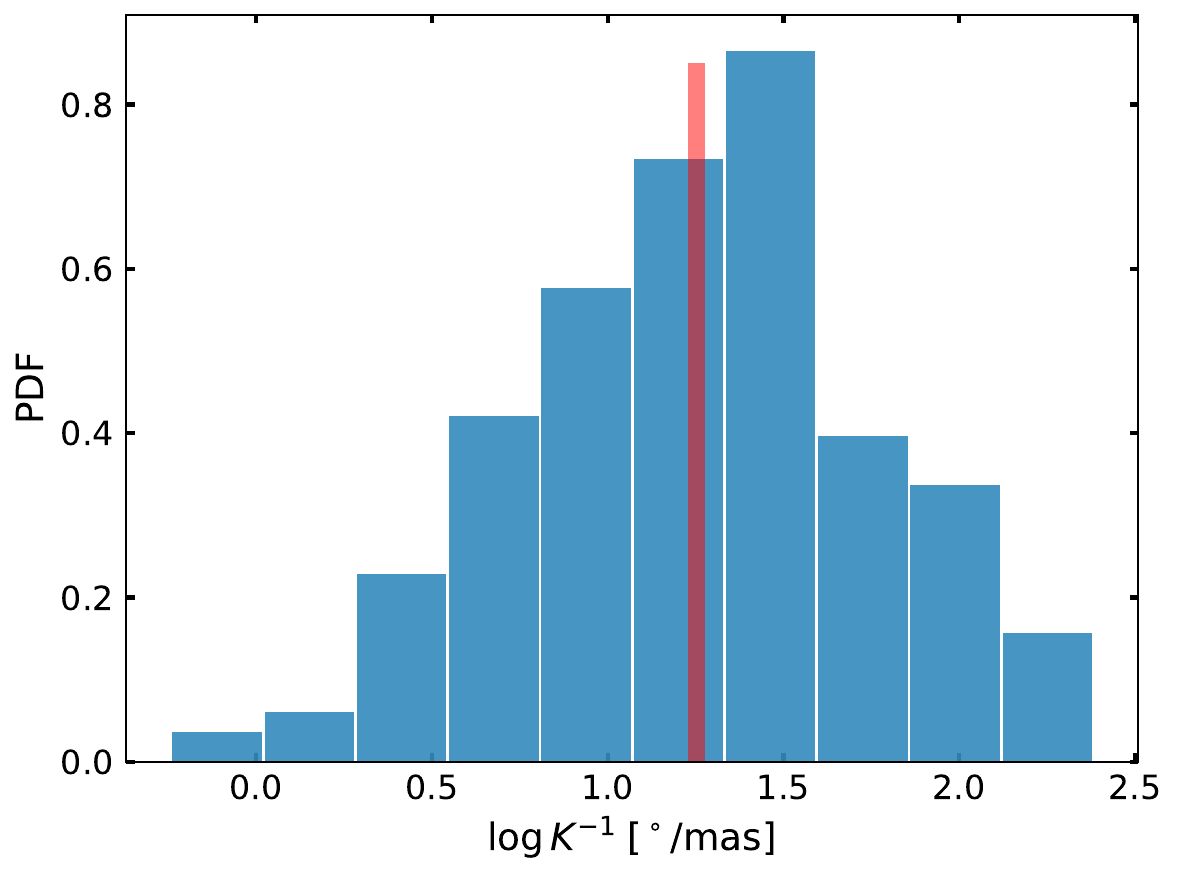}\\[0.3em]
            {\small (b) Linear helix model ($\alpha > 0$)}
            \label{fig:K_M1_lin_b}
        \end{minipage}
        \hfill
        \begin{minipage}[t]{0.32\linewidth}
            \centering
            \includegraphics[width=\linewidth]{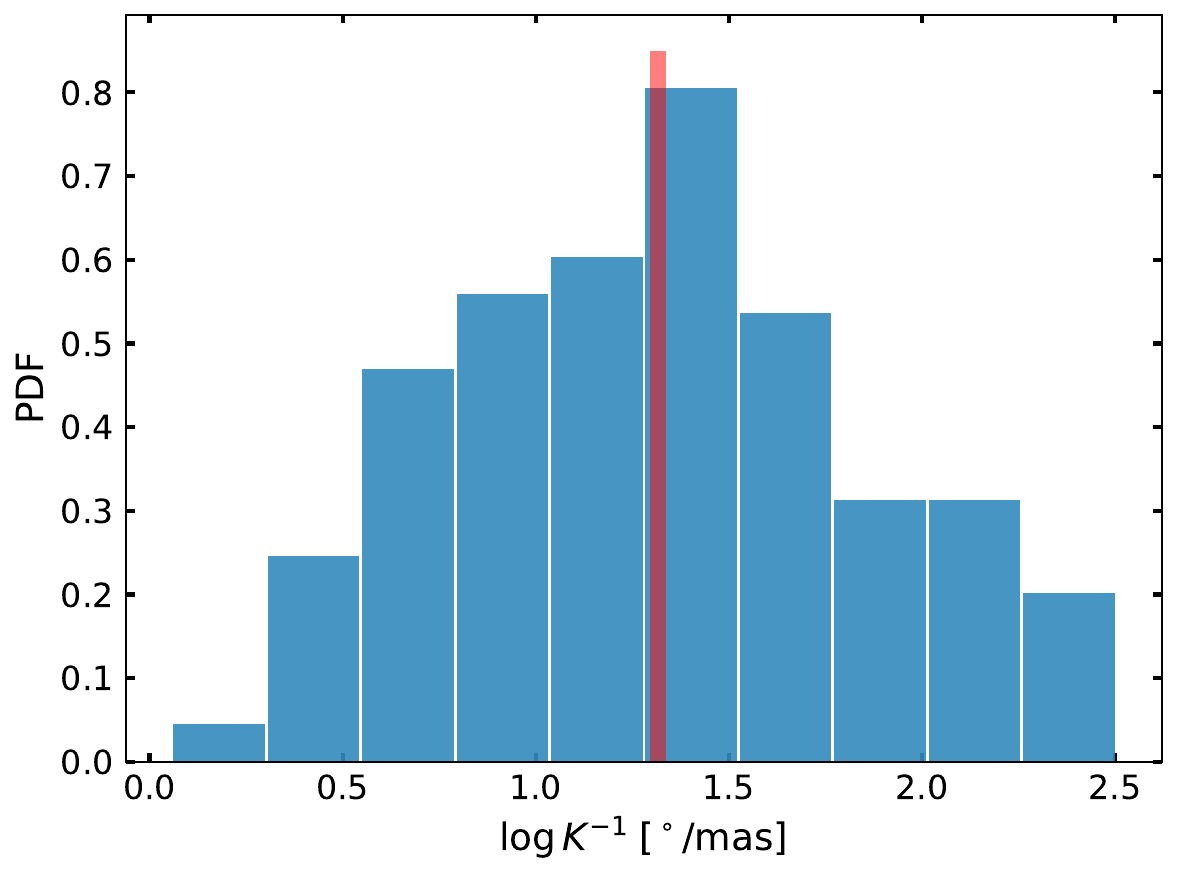}\\[0.3em]
            {\small (c) Best linear helix models ($\alpha \geq 0$)}
            \label{fig:K_Mpref_lin_c}
        \end{minipage}
        
        \vspace{1em}

        \begin{minipage}[t]{0.32\linewidth}
            \centering
            \includegraphics[width=\linewidth]{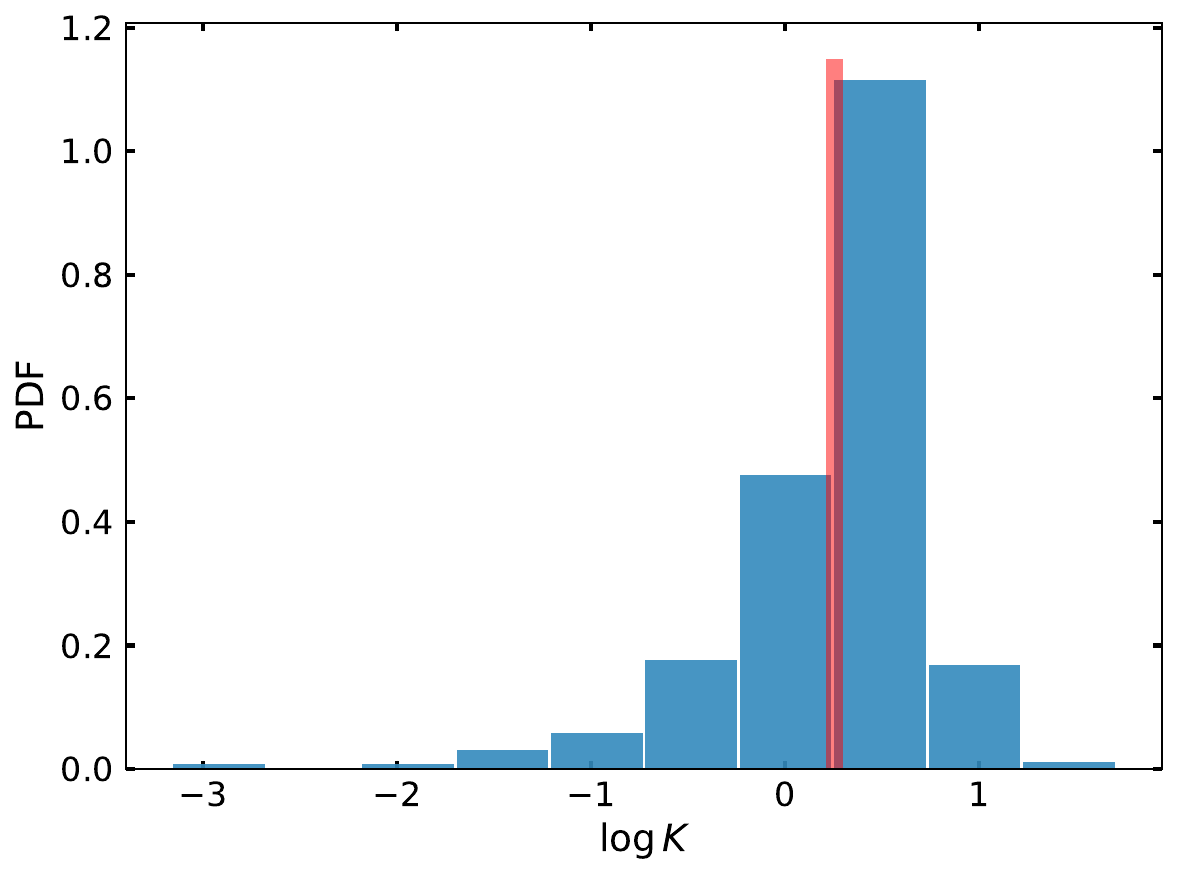}\\[0.3em]
            {\small (d) Logarithmic helix model ($\alpha = 0$)}
            \label{fig:K_M0_log_d}
        \end{minipage}
        \hfill
        \begin{minipage}[t]{0.32\linewidth}
            \centering
            \includegraphics[width=\linewidth]{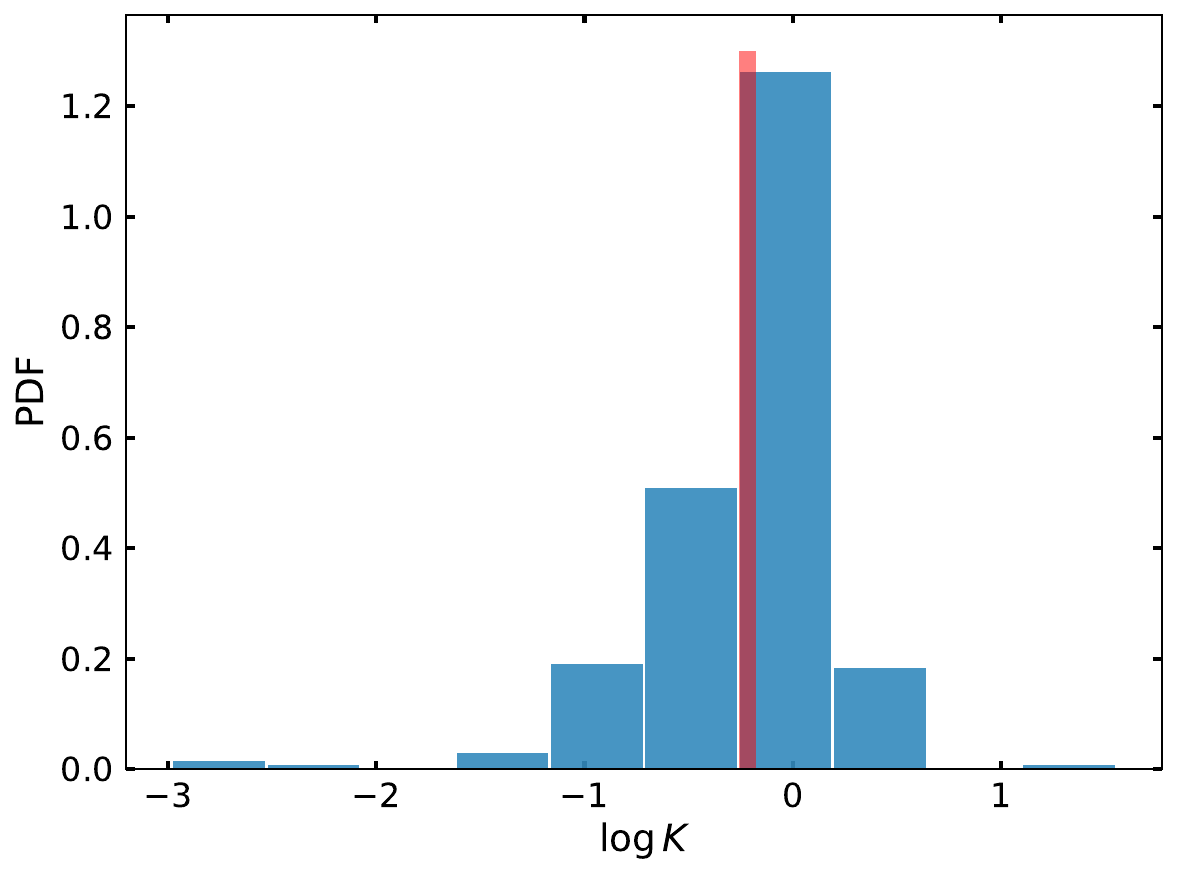}\\[0.3em]
            {\small (e) Logarithmic helix model ($\alpha > 0$)}
            \label{fig:K_M1_log_e}
        \end{minipage}
        \hfill
        \begin{minipage}[t]{0.32\linewidth}
            \centering
            \includegraphics[width=\linewidth]{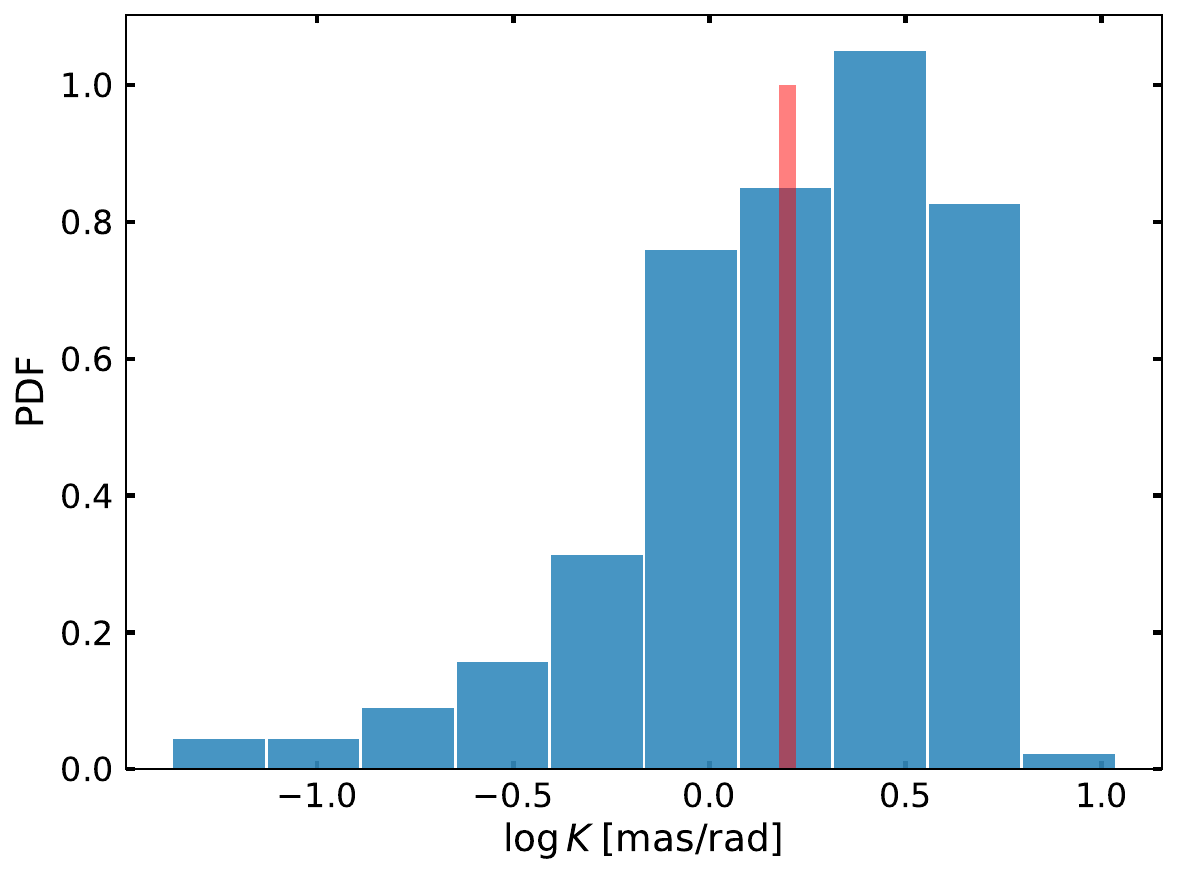}\\[0.3em]
            {\small (f) Best logarithmic helix models ($\alpha \geq 0$)}
            \label{fig:K_Mpref_log_f}
        \end{minipage}
        
        \caption{Distribution of the parameter $K$ based on the best-fit results for the various tested models, together with the distribution of $K$ for the subset of selected models.  The logarithmic mean of each distribution is indicated by a vertical red line. The type of each model is indicated below its respective panel.}
        \label{fig:K_fits}
        \end{figure*}
        
        \subsection{Fitting results}
        
            The final fitting results for the linear helix model for both $\alpha = 0$ and $\alpha > 0$ are presented in Table~\ref{tab:jet_hel_fits}. In the case of $\alpha > 0$, for a subset of sources, the Hessian at the optimum was not positive definite, indicating locally flat or saddle-shaped metric surfaces and hence a lack of identifiability for one or more parameters. For these sources, we report only the point estimates. These cases are excluded from further statistical analysis. The same applies to the logarithmic helix model, the optimal parameters of which are provided in Table~\ref{tab:jet_loghel_fits}.
            
            For further analysis, to ensure fit robustness, we exclude fits with $\chi^2_{\rm red}>10$ or fewer than 40 ridge-line points ($N<40$). Out of the total sample of 586 sources, the resulting linear-helix fits comprise 512 sources for $\alpha=0$ and 317 for $\alpha\gg0$, with 306 sources in common between the two cases. For the logarithmic-helix model, the corresponding numbers are 498, 288, and 280 sources, respectively.

            \begin{figure}
                \centering
                \includegraphics[width=1\linewidth]{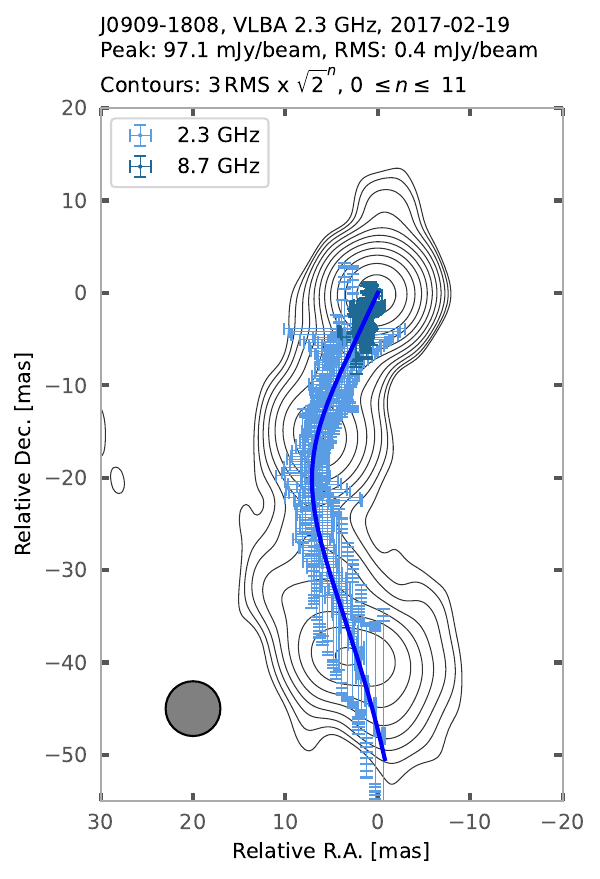}
                \caption{Example of the best-fit model (blue line) for the ridge lines of the jet of the quasar J1909$-$1808. The ridge lines were constructed using 2.3~GHz and 8.7~GHz data and are shown on the plot with error bars in purple and pale blue, respectively. The image of the quasar from Fig.~\ref{fig:curved_example} is shown in the background. The FWHM of the restoring beam is shown in the bottom left corner as a shaded circle.}
                \label{fig:best_fit_example}
            \end{figure}

            The first important observable description of our sample is the value of $K^{-1}$ for the linear helix fit with $\alpha = 0$. That is the simplest model, and it describes overall bending trends among curved jets in our sample. The distribution of that parameter is shown in Fig.~\ref{fig:K_fits}(a). The average value of $\log{K^{-1}}$ is 2.04, which corresponds to curvature $K^{-1}_{\text{mean}} = 109^\circ\!/ \text{mas}$. On a typical scale of $10 \text{ mas}$ this results in $1090^\circ$ apparent twist, which is not observed in our sample. Given that the number of observations at longer wavelengths is lower than at shorter wavelengths, we propose two possible scenarios explaining that discrepancy. First, the actual values of $\alpha \gg 0$ for most sources in the sample; second, the helix wavelength is increasing with distance from the core, which favours the model of plasma instability. Both options are supported by the observed dependence of the bending rate on frequency (Fig.~\ref{fig:freq_bending}).
            
            The distributions of $\log K$ for the other model variations are shown in panels (b), (d), and (e) of Fig.~\ref{fig:K_fits}. Their separate analysis is no longer as descriptive as for the linear helix with $\alpha = 0$ and is therefore conducted only after the model selection procedure described further.
            
        \subsection{Model selection}

            To select models from our list that best fit the data, we perform a comparative analysis based on the Bayesian information criterion (BIC) \citep{Liddle2007}. For a model with $k$ free parameters fitted to $N$ points using the $\chi^2$ method, $\text{BIC} = \chi^2 + k\ln{N}$. By comparing values of $\text{BIC}$ one can understand which model describes the data better while also being balanced by the number of parameters. The lower the BIC, the better the model. A difference in BIC between two models of $|\Delta\text{BIC}| \lesssim 4$ indicates similar descriptive power, $4 \lesssim |\Delta\text{BIC}| \lesssim 10$ indicates a potential advantage of one model over another, and $|\Delta\text{BIC}| > 10$ provides strong evidence that one model describes the data better than another. We choose the last option to ensure robustness of the final selection. In addition, we assume that the fit for $\alpha = 0$ is automatically better than the one with $\alpha \gg 0$ if the estimate of $\alpha$ is weakly constrained ($\alpha < 3\sigma_\alpha$). Except in the last case, fits with $|\Delta\text{BIC}| < 10$ are not included in the final statistics.

            We first compare the two models within each helix family and select the preferred one. We then compare the preferred linear and logarithmic helix models. A source is included only if both fits are valid. Within the linear-helix family, the initial model selection favours 16 fits with $\alpha=0$ and 167 with $\alpha>0$, while for the logarithmic-helix family it favours 147 and 39 fits, respectively. After the final comparison between the two model families, only 43 models remain in the linear case, of which 3 have $\alpha=0$ and 40 have $\alpha>0$, whereas only 8 logarithmic models remain, all with $\alpha=0$. This is not enough for a reliable physical conclusion. Moreover, the observed imbalance between the final models may be partly caused by the greater numerical complexity of the logarithmic model for $\alpha>0$ compared to the others, which results in lower fit quality than for the logarithmic fit with $\alpha=0$ and for the general linear fit. Furthermore, if the change in the jet radius is negligible, we might not be able to detect the change in $\lambda$ (as in Eq.~\ref{eq:Hardee_int}), and the final helix would be approximated well by a linear helix. These observations, however, do not prevent us from analysing both logarithmic helix fits separately from the linear helix fit under the assumption that the models cannot be robustly differentiated.

            Separately from the logarithmic helix, using selected best-fit linear helix models, we establish that 78\% of sources have $\alpha > 1$, 21\% have $\alpha < 1$, 42\% of which have $\alpha = 0$ (or more generally $\alpha \sim 0$). Under the assumption of the jet precession model, $\alpha > 1$ means that the precession cone is observed from the outside, whereas if $\alpha < 1$, we are looking inside the precession cone. The median value of $\alpha$ equals $2.55$. The distribution of parameter $K$ is shown in Fig.~\ref{fig:K_fits}(c). The log-mean value corresponds to the curvature $K^{-1}_{\text{mean}} = 20^\circ / \text{mas}$, which would give $\sim200^\circ$ intrinsic twist at the 10~mas separation from the core.

            For the logarithmic helix, the $\alpha = 0$ case is favoured for most sources. The final distribution of $K$ is shown in Fig.~\ref{fig:K_fits}(f). The log-mean value is $K_{\text{mean}} = 1.63 \text{ rad} = 93^\circ$, which means that if the distance from the core increases by a factor of two, the twist would be $\Delta \varphi = K\ln2 = 186^\circ$. However, due to fitting limitations, we might expect more sources that actually have $\alpha > 0$, which shifts the distribution of $K$ towards a higher mean value and more twisted helices, as observed from the difference between Fig.~\ref{fig:K_fits}(d) and Fig.~\ref{fig:K_fits}(e).
            
        \subsection{Estimates of precession periods}
            
\begin{table*}
    \centering
    \caption{Fitted parameters for the linear helix models. Subscripts 0 and 1 denote the models with $\alpha=0$ and $\alpha>0$, respectively.}
    \label{tab:jet_hel_fits}
    
    \footnotesize
\setlength{\tabcolsep}{4pt}
    
    \begin{tabular}{lccccccccc}
        \hline
        \hline
        Source
        & $\mathrm{PA}_0$
        & $K_0$
        & $\chi_{\mathrm{red},0}^2$
        & $\mathrm{PA}_1$
        & $\Theta_1$
        & $\alpha_1$
        & $K_1$
        & $s_1$
        & $\chi_{\mathrm{red},1}^2$ \\
        & & (mas) & & & & & (mas) & & \\
        \hline

        J0001$-$1551
        & $-0.0298\pm0.0019$
        & $-1.587\pm0.011$
        & 0.12
        & $0.984\pm0.022$
        & $2.444\pm0.273$
        & $3.425\pm0.213$
        & $1.775\pm0.259$
        & $1$
        & 0.076 \\

        J0008$+$1144
        & $0.0096\pm0.0021$
        & $0.52\pm0.19$
        & 0.06
        & $-2.528$
        & $-1.954$
        & $0.943$
        & $29.485$
        & $1$
        & 0.047 \\

        J0008$+$6837
        & $0.023\pm0.005$
        & $0.2\pm1.4$
        & 1.36
        & $1.371\pm0.038$
        & $-0.199\pm0.120$
        & $2.824\pm0.063$
        & $1.657\pm0.039$
        & $1$
        & 0.978 \\

        J0019$+$7327
        & $0.2244\pm0.0006$
        & $0.04\pm7.31$
        & 2.38
        & $-2.863\pm0.009$
        & $-2.463\pm0.079$
        & $3.994\pm0.077$
        & $0.503\pm0.021$
        & $-1$
        & 0.449 \\

        J0023$+$0101
        & $0.13\pm0.07$
        & $0.7\pm0.8$
        & 0.04
        & $2.343$
        & $-0.319$
        & $0.325$
        & $9.565$
        & $1$
        & 0.022 \\

        J0024$+$4644
        & $-0.0092\pm0.0015$
        & $0.32\pm0.25$
        & 0.06
        & $-2.377\pm0.641$
        & $1.657\pm1.147$
        & $1.078\pm0.673$
        & $29.919\pm26.917$
        & $-1$
        & 0.053 \\

        \hline
        \multicolumn{10}{p{0.93\textwidth}}{\footnotesize
        Columns are as follows: (1) source name (J2000); parameters with subscripts 0 and 1 correspond to the $\alpha=0$ and $\alpha>0$ models, respectively. See Section~\ref{sec:geomodel} for definitions of the model parameters. The full table is available electronically from CDS; the first six entries are shown here for reference.}
    \end{tabular}
\end{table*}

\begin{table*}
    \centering
    \caption{Fitted parameters for the logarithmic helix models. Subscripts 0 and 1 denote the models with $\alpha=0$ and $\alpha>0$, respectively.}
    \label{tab:jet_loghel_fits}

    \footnotesize
    \setlength{\tabcolsep}{4pt}

    \begin{tabular}{lccccccccc}
        \hline
        \hline
        Source
        & $\mathrm{PA}_0$
        & $K_{\log, 0}$
        & $\chi_{\mathrm{red},0}^2$
        & $\mathrm{PA}_1$
        & $\Theta_1$
        & $\alpha_1$
        & $K_{\log, 1}$
        & $s_1$
        & $\chi_{\mathrm{red},1}^2$ \\
        & & (mas) & & & & & (mas) & & \\
        \hline

        J0001$-$1551
        & $-0.314\pm0.020$
        & $-0.295\pm0.048$
        & 0.125
        & $3.142$
        & $3.142$
        & $1.000$
        & $47.348$
        & $-1$
        & 9.950 \\

        J0008$+$1144
        & $0.156\pm0.033$
        & $1.671\pm0.096$
        & 0.057
        & $-2.888$
        & $-0.034$
        & $1.162$
        & $0.553$
        & $1$
        & 0.044 \\

        J0008$+$6837
        & $0.277\pm0.043$
        & $4.306\pm0.107$
        & 1.368
        & $0.765\pm0.187$
        & $-0.388\pm0.240$
        & $0.823\pm0.162$
        & $0.633\pm0.180$
        & $1$
        & 1.023 \\

        J0019$+$7327
        & $7.344\pm0.009$
        & $15.710\pm0.018$
        & 6.799
        & $-2.402\pm0.031$
        & $-0.168\pm0.018$
        & $1.402\pm0.033$
        & $0.180\pm0.020$
        & $-1$
        & 0.476 \\

        J0023$+$0101
        & $0.703\pm0.375$
        & $0.942\pm0.666$
        & 0.035
        & $-2.588\pm0.574$
        & $-0.277\pm0.429$
        & $1.102\pm0.481$
        & $1.688\pm1.415$
        & $1$
        & 0.029 \\

        J0024$+$4644
        & $-0.133\pm0.020$
        & $3.281\pm0.049$
        & 0.049
        & $2.621\pm0.124$
        & $0.079\pm0.097$
        & $1.393\pm0.161$
        & $0.189\pm0.068$
        & $1$
        & 0.056 \\

        \hline
        \multicolumn{10}{p{0.93\textwidth}}{\footnotesize
        Columns are as follows: (1) source name (J2000); parameters with subscripts 0 and 1 correspond to the $\alpha=0$ and $\alpha>0$ models, respectively. The subscript $\log$ denotes the logarithmic-helix parameter $K$. See Section~\ref{sec:geomodel} for definitions of the model parameters. The full table is available electronically from CDS; the first six entries are shown here for reference.}
    \end{tabular}
\end{table*}

\begin{figure}
                        \centering
                        \includegraphics[width=\linewidth]{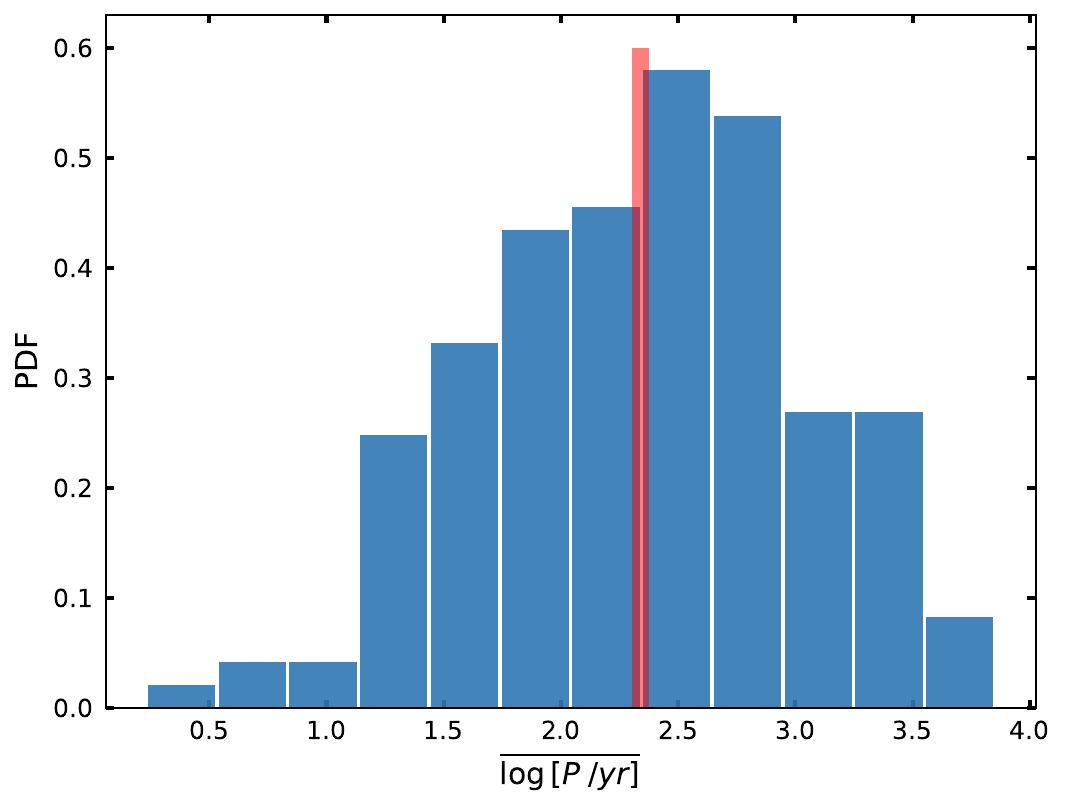}
                        \caption{Distribution of precession periods derived from a Monte Carlo analysis of the curvature parameter $K^{-1}$ obtained from the fitted linear helix model. The logarithmic mean of the distribution is indicated by the vertical red line.}
                        \label{fig:P_dist}
                    \end{figure}

            Using the Monte-Carlo modelling, we obtain the posterior distributions for the precession period (its logarithm) $\log P$ for each source. From each posterior we establish the mean value $\overline{\log P}$ and set up a new distribution which has a probability density function shown in Fig.~\ref{fig:P_dist}. Overall, the periods range from $P \sim 10$ to $10^{4}$ years, with a logarithmic mean of $218.8$ years, which is consistent with estimates provided by many other authors when analysing individual objects \citep{precbh, smbhb, beam_prec, Lobanov2005}.

        \subsection{Supermassive black hole binary model}
            We consider the case where we have a system of two SMBHs in an active galactic nucleus. The mass ratio between them is given by $ q = \frac{m}{M} < 1 $, where $ M $ and $ m $ are the masses of the primary and secondary SMBHs. We also assume that the primary SMBH is surrounded by an accretion disc that fuels the jet. In such a system, the orbit of the less massive black hole (BH) can be considered circular and intersects the accretion disc at $ r_{b} < r_{d} $, where $ r_b $ is the orbital radius and $ r_d $ is the disc’s outer radius, but the orbit and the disc are not coplanar. Theory suggests that if a jet exists in such an SMBHB, its base might undergo precession, and there are several mechanisms that can describe that phenomenon. The first one is driven by the spin geodetic precession of the primary SMBH due to the interaction with the secondary. The period of this precession, as shown in \citet{precbh}, can be estimated as:
             \begin{equation} \label{eq:geod_prec_full_}
                P_{\text{geo}} \approx 2.6 \cdot 10^8 M_8 \,q^{-1} \left( \frac{r_b}{10^4 r_g}\right)^{5/2} \text{ [yr] },
            \end{equation} 
            where $M_8 = M \cdot 10^{-8}$ and $r_g = 2 G M  /c^{2}$ -- gravitational radius of the primary SMBH.

            The second precession mechanism in our consideration is based on the precession of the disc due to the secondary SMBH. If the orbit of the secondary lies outside the accretion disc, the latter may undergo solid-body precession \citep{Larwood1997} (a similar expression is also derived by \citet{Katz1997}, differing by a factor of $\sim 2$), which might cause precession of the jet. The period is estimated as
            \begin{equation} \label{eq:P_td_0}
                P_{\text{td}} \approx 6.5 \cdot 10^{2} \, M_8 \, q^{-1} 
                \left( \frac{r_b}{10^4 r_g} \right)^3 
                \left( \frac{R_d}{10^4 r_g} \right)^{-\frac{3}{2}}  
                \text{ [yr] },
            \end{equation}
            where $R_d < r_b$ is the disc size. Given this constraint, we derive a lower boundary for the precession period with the disc size remaining unknown:
            \begin{equation} \label{eq:P_td}
                P_{\text{td}} \gtrsim 6.5 \cdot 10^{2} \, M_8 \, q^{-1}
                \left( \frac{r_b}{10^{4} r_g} \right)^{\frac{3}{2}} 
                \text{ [yr] }.
            \end{equation}
            From Eq.~\ref{eq:geod_prec_full_} and Eq.~\ref{eq:P_td_0} we immediately see that $P_\text{td} \ll P_\text{geo}$: 
            \begin{equation}
                \frac{P_{\text{td}}}{P_{\text{geo}}} = 2.6 \cdot{10^{-6}}  \left(\frac{r_b}{10^4r_g}\right)^{\frac{1}{2}} \left(\frac{R_d}{10^4r_g}\right)^{-\frac{3}{2}}. 
            \end{equation}
                
            We study precession under the assumption that the jet structure does not change during the observational span and can be physically described as a linear helix with a fixed precession period. In the binary system, the main parameter that changes during the evolution is the separation $r_b$ of the SMBHs. The timescale on which this change becomes significant is described by the merger time $t_{\text{merge}}$. Therefore, we may interpret $P_{\text{obs}} \approx \text{const}$ as the condition $t_{\text{merge}} \gg P_{\text{obs}}$. The merger time, as given in \citet{merge}, can be estimated as
            \begin{equation}
            \begin{aligned}
                t_{\text{merge}}  \approx   2.9 \cdot 10^{12} \cdot  \left( \frac{m}{10^6 \cdot M_{\odot}} \right)^{-1} \left( \frac{M}{10^6 \cdot M_{\odot}} \right)^{-1} \\ \times  \left( \frac{m + M}{2 \cdot 10^6 \cdot M_{\odot}} \right)^{-1} \left( \frac{r_b}{10^{-2} \text{pc}} \right)^{4}  \text{ [yr] },
            \end{aligned}
            \end{equation}
            which can be rewritten in terms of mass ratio $q$ and primary mass $M_8$ as
            \begin{equation} \label{eq:merge}
                t_{\text{merge}} \approx 4.9 \cdot 10^{12} M_8 \,q^{-1}(1+q)^{-1} \left(\frac{r_b}{10^4 r_g}\right)^4 \text{ [yr] },
            \end{equation}
            where the factor $(1+q)^{-1}$ can potentially be neglected for order-of-magnitude estimates since its contribution to the equation is very small (a factor of $0.5 \div 1$). Using Eq.~\ref{eq:geod_prec_full_} and Eq.~\ref{eq:merge}, we convert the inequality connecting the precession period and the merger time into a lower limit for the separation between SMBHs, $r_b \gg 10\, r_g \sim 10^{-4} \text{ pc}$.
            
            A SMBHB with $q \sim 1$ would produce double jets, which are not seen in our sample \citep{Moesta2012, Palenzuela2010}. Furthermore, if we assume a short separation, the jet becomes driven not by the spin of the primary BH but by the total angular momentum of the system \citep{Palenzuela2010}, which does not precess. Therefore, cases with small separations $r_b \sim 10 r_g$ and black holes of comparable masses $q \sim 1$ may be considered less likely to occur. 
            
            Additionally, \citet{Ellison2019} showed that AGN in major mergers typically exhibit observable merger signatures. Hence, the dominant mechanism for SMBHB formation is the migration of the secondary SMBH into the centre of a more massive galaxy after a merger event. This implies that the SMBH of the less massive galaxy must “sink’’ towards the primary SMBH due to dynamical friction. As shown by \citet{Dosopoulou2017}, for SMBHs with mass ratios $q \lesssim 10^{-3}$ the timescale of this process exceeds a Hubble time, and the SMBHs will not approach each other closely enough (their separation will remain larger than the scale of the primary's accretion disc). This sets a lower boundary of $q \gg 10^{-3}$ for our SMBHB case.

\begin{figure}
                
                \includegraphics[width=\linewidth]{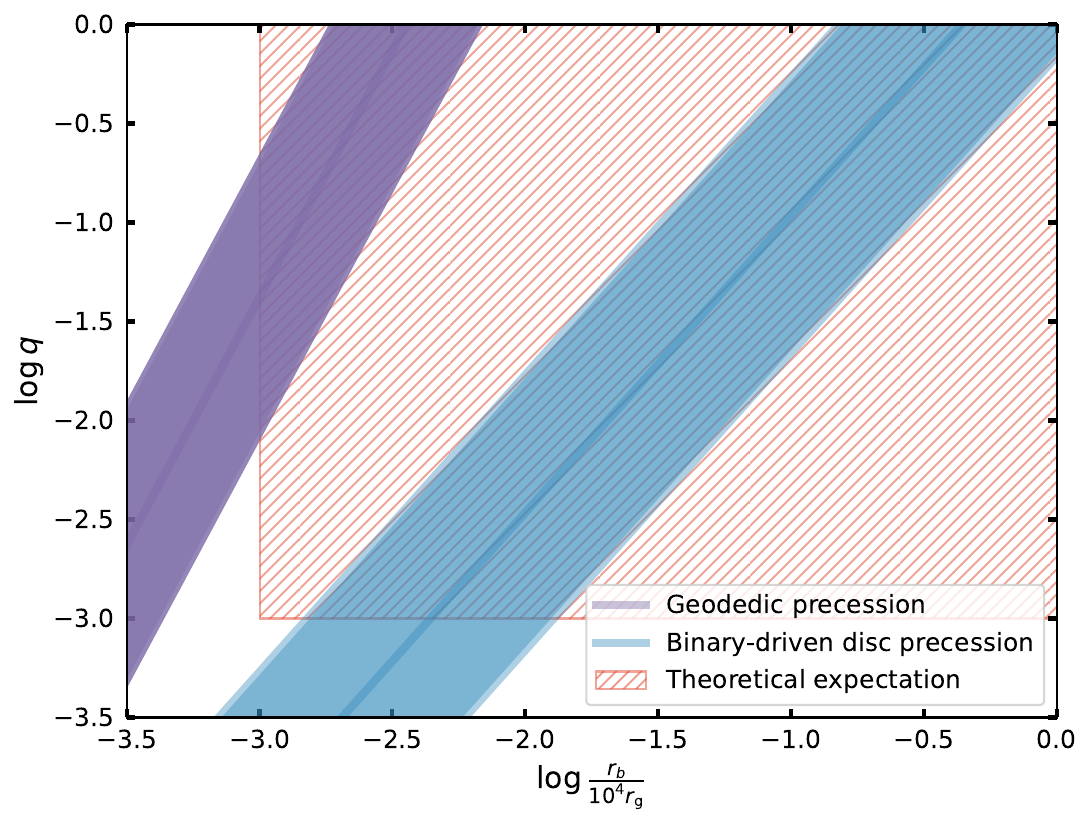}
                \caption{Precession in models involving SMBHBs: the plot shows the allowed combinations of black hole mass ratio and separation. The blue and purple shaded regions represent the allowed values of the SMBHB separation $r_b$ and mass ratio $q$, derived by combining the mean and dispersion of the precession periods estimated in this work with the mean SMBH mass and its dispersion for quasars reported by \citet{mass} (both within $\pm\sigma$ and $\pm2\sigma$). The hatched reddish area marks the region of $(q, r_b)$ parameter space favoured by theoretical considerations.}
                \label{fig:mratio_sep_plot}
            \end{figure}

            In order to investigate the parameter space of the considered precession models and compare it with theoretical limitations, we study which parameters are allowed for precession periods within $\overline{\log P_\text{obs}} \pm \sigma_{\log P}$ and $\overline{\log M} \pm \sigma_{\log M}$, where for the mass distribution we adopt the AGN mass function presented by \citet{mass}. The results are shown in Fig.~\ref{fig:mratio_sep_plot}. 

            For the geodetic precession model, the figure shows that under the given model constraints, all predicted corresponding pairs of separation and mass ratio fall within a very narrow range of tightly bound SMBHs with similar masses ($r_b \sim (10 \div 30)\, r_g$ and $q \gtrsim 10^{-1}$). In contrast, the model of solid-body disc precession satisfies the constraints over a wide parameter range in both $r_b$ and $q$. However, it is important to note that since we are working with the lower boundary for which $R_d \sim r_b$, the physically valid values of $r_b$ for this model may be further constrained by the accretion disc size, $r_g \lesssim R_d \, 10^{-(3 \div 2)}$ \citep{McHardy2023, Goodman2003}. This constraint still allows a large range of possible $r_b$, but connects the estimated periods to higher values of the mass ratio ($q \gtrsim 10^{-2}$).
         
        \subsection{Tilted accretion disc model}
        
            Another mechanism for the precession of the AGN jet can be Lense--Thirring precession. This phenomenon is predicted within the framework of general relativity and is associated with the influence of the gravitational field of a rotating body on the motion of other bodies in its vicinity. This effect manifests itself as an additional rotation (precession) of the body's orbit or its angular momentum vector. As described in \citet{lentir}, the outer regions of the accretion disc themselves can cause Lense--Thirring precession of the central compact object and the inner part of the accretion disc. The angular momentum precession of the central black hole determines the base precession of the jet. In the $\alpha$-disc approximation, as shown by \citet{Martin2007, Perego2009, Fiacconi2018, Dotti2013}, the period of such precession is of the order of the alignment timescale of the jet and disc spins and can be expressed as:
            \begin{equation} 
            \label{eq:aprec}
            P_{\text{BH, spin}} \approx 3.9 \cdot 10^{6} f_{\nu_2}^{-\frac{5}{7}} s^{\frac{5}{7}} \alpha^{\frac{58}{35}}  M_8^{-\frac{2}{35}} \left(\frac{\dot{m}}{ \eta_{0.1} }\right)^{-\frac{32}{35}} \text{ [yr] },
            \end{equation}
            where $s$ is the spin parameter of the black hole, $\dot{m}$ is the accretion rate of the black hole, $\eta_{0.1} = \eta / 0.1$ is the accretion efficiency, and $\alpha$ is the disc viscosity. The term $f_{\nu_2}$ describes the relation between the usual viscosity coefficient, $\nu_1 = \alpha c_s H$ (where $c_s$ is the sound speed in the disc and $H$ is the disc scale height), and the disc warp diffusion coefficient, $\nu_2$. These quantities are related as follows \citep{Lodato2007}:
            \begin{equation}
            \frac{\nu_2}{\nu_1} = \frac{f_{\nu_2}}{2 \alpha^2}.  
            \end{equation}
            However, given that the value of $f_{\nu_2}$ does not change significantly even if $\alpha$ is varied by an order of magnitude \citep{Lodato2007}, we assume $f_{\nu_2} \sim 1$ throughout. Under this assumption, the final expression for the precession period is
            \begin{equation}\label{eq:prec_disk}
            P_{\text{BH, spin}} \approx 3.9 \cdot 10^{6} s^{\frac{5}{7}} \alpha^{\frac{58}{35}}  M_8^{-\frac{2}{35}} \left(\frac{\dot{m}}{ \eta_{0.1} }\right)^{-\frac{32}{35}} \text{ [yr] }.
            \end{equation}
            
            If the jet is also connected to the spin of the accretion disc, there may be another mechanism that can cause precession, namely warping of the inner region of the accretion disc \citep{Dotti2013}. The warping timescale, with the aforementioned assumptions on $f_{\nu_2}$, is expressed as
            \begin{equation} \label{eq:prec_inner}
            P_\text{inner disc} \approx 8.6 \cdot s^{\frac{5}{7}}\alpha^{\frac{72}{35}}M_8^{\frac{47}{35}}\left(\frac{\dot{m}}{ \eta_{0.1} }\right)^{-\frac{18}{35}} \text{ [yr] }.
            \end{equation}
            
            In practice, thin $\alpha$-discs have several restrictions. It is theoretically predicted that for $\dot{m} \lesssim \alpha^2$ the disc may transition into an advection-dominated accretion flow (ADAF) regime and will no longer be thin \citep{Narayan1994, Narayan1995, Qiao2009}. In addition, for higher values of the accretion rate ($\dot{m} \gtrsim 1$) the disc may become slim, in which case the thin $\alpha$-disc approximation also breaks down \citep{Abramowicz1988, Mineshige2000, Sadowski2009}. These constraints are purely theoretical and indicate the existence of additional solutions that may dominate; therefore, in our study we consider them in a relaxed form, allowing for $1\div2$ orders of magnitude deviation.
            
            From an observational perspective, the bolometric luminosity of a source traces the actual accretion rate: $L_\text{bol}/L_\text{Edd} = \eta_\text{rad}\dot{m}$, where $L_\text{Edd}$ is the Eddington luminosity and $\eta_\text{rad}\sim10^{-(1\div3)}$ is the radiative efficiency. Therefore, we may introduce an additional observational constraint on the accretion rate, roughly $-3\lesssim\log\dot{m} \lesssim 2\div3$ \citep{Koss2022, Koss2022cat}. In addition, the accretion disc is not expected to have extremely low or high viscosity values: $0.001 \lesssim \alpha \lesssim 0.4$ \citep{Siemiginowska1989, Starling2004, King2007}.
            
\begin{figure}
                \centering
                \includegraphics[width=0.999\linewidth]{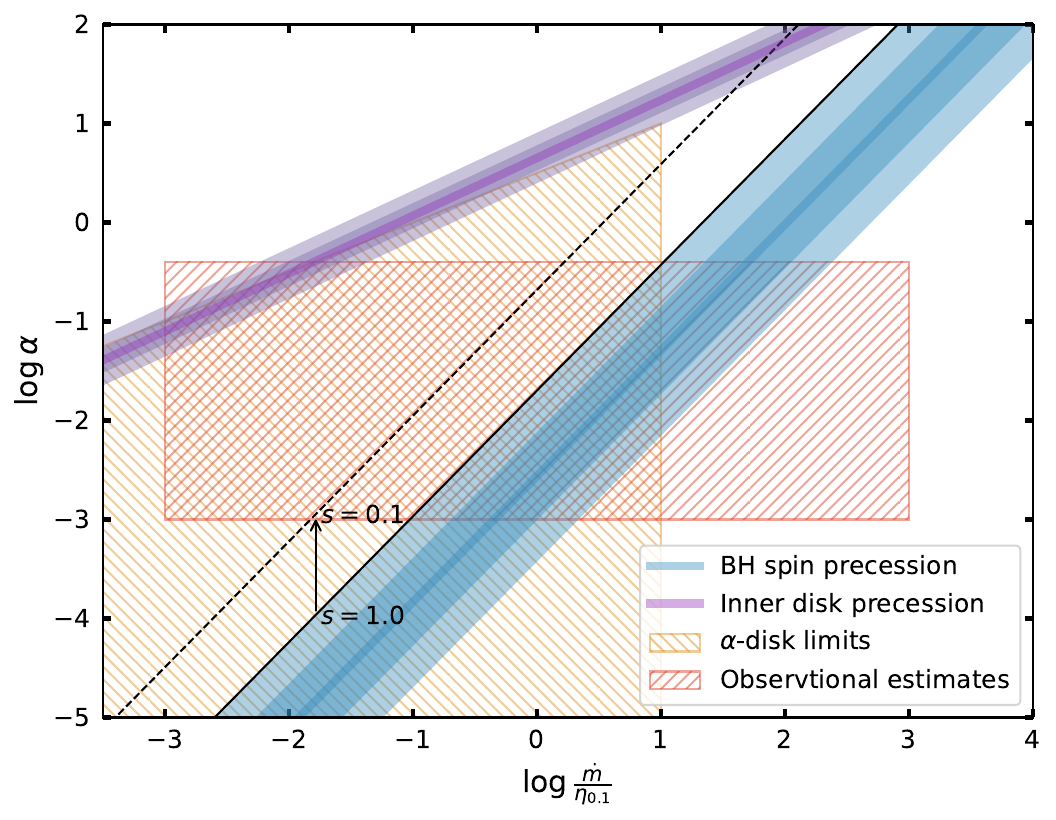}
                \caption{Precession in the Lense--Thirring model: a plot of the allowed values of viscosity and accretion rate (in Eddington units). The blue and purple shaded regions represent the allowed values of the viscosity $\alpha$ and mass accretion rate in Eddington units $\dot{m}/\eta_{0.1}$, derived by combining the mean and dispersion of the precession periods estimated in this work with the mean SMBH mass and its dispersion for quasars reported by \citet{mass} (both within $\pm\sigma$ and $\pm2\sigma$). The arrow indicates where the $+2\sigma$ boundary for the inner disc precession model moves if spin is decreased from 1 to 0.1 (black solid line $\rightarrow$ black dashed line). Hatched red and yellow areas show values of $\alpha$ and $\dot{m}$ that are allowed by observations and theory.}
                \label{fig:visc_ar_plot}
            \end{figure}

            We utilise Eq.~\ref{eq:prec_disk} and Eq.~\ref{eq:prec_inner} in the same way as for the SMBHB model, taking into account the discussed theoretical and observational limits. The disc viscosity $\alpha$ and the accretion rate $(\dot{m}/\eta_{0.1})$ are used as free parameters. The spin parameter of the black hole is set to unity, $s = 1$, for simplicity, since all estimates are made at the order-of-magnitude level. To account for the effect of spin, we also show how the results change if the spin is reduced to $s = 0.1$. The allowed parameter combinations are presented in Fig.~\ref{fig:visc_ar_plot}. We conclude that the model of jet precession originating from inner disc warping is valid only in a very narrow region of low accretion rate and high viscosity. Theoretically, this region lies close to the ADAF boundary, which makes it less favourable than precession driven by SMBH and disc spin alignment timescales. The latter model, assuming $\eta_{0.1} \sim 1$, favours Eddington and mildly super-Eddington accretion rates, $\dot{m} \sim 0.02 \div 10$. The lower limit corresponds to the lower spin estimate ($s = 0.1$).
            
            So far we have discussed only thin discs; however, jet precession arising from SMBH spin Lense--Thirring precession is also possible for a thick, tilted disc \citep{Fragile2007, Stone2012}. In that case, the main body of the accretion disc precesses relative to the SMBH spin, which is kinematically similar to the precession of the SMBH spin about the angular-momentum vector of the accretion disc. The period of such precession can be roughly estimated as
            \begin{equation}
            P_\text{BH, spin, thick} \approx 24.7 \cdot s^{-1} M_8 \left(\frac{r_\text{o}}{100r_g}\right)^3 \left(\frac{r_\text{o}}{100 r_\text{i}}\right)^{\frac{1}{2}} \text{[yr]},
            \end{equation}
            where $r_\text{i}$ and $r_\text{o}$ are the inner and outer radii of the disc. The order of magnitude of the precession period predicted by this model for $r_\text{o} \sim 10^{1\div3}\,r_g$ is $P_\text{BH, spin, thick} \sim 10^{1\div5}$~yr, which is consistent with our period estimates and makes this model favourable for explaining jet precession in the case of a thick disc.

        \subsection{Helical instability model}
        
           In addition to precession within the framework of the linear helix model, we also consider the effect of the Kelvin--Helmholtz instability. The dependence for the logarithmic spiral morphology can be rewritten in the form of Eq.~(\ref{eq:KH_shape}), where the coefficient in front of the logarithm is
            \begin{equation}
                K_{\text{log}} = \frac{2 \pi}{\tan{\psi}} \frac{r_0}{\lambda_0},
            \end{equation}
            from which
            \begin{equation} \label{eq:lambda_ratio_param}
                \frac{r_0}{\lambda_0} = \frac{2 \pi K_{\text{log}}}{\tan{\psi}} .
            \end{equation}
            
            Theoretical analysis by \citet{KH} suggests that for the resonant instability mode under consideration $\lambda_0 / r_0 \sim \gamma \mathcal{M}_\text{in}$, where $\gamma$ and $\mathcal{M}_\text{in}$ are the bulk Lorentz factor and the Mach number of the plasma in the jet, respectively. Therefore, given that typical values are $\gamma \sim 5 \div 15$ and the motion is supersonic, the following condition must be satisfied:
            \begin{equation} \label{eq:lambda_ratio_crit}
                \frac{r_0}{\lambda_0} \ll 1 .
            \end{equation}
            This condition can be interpreted as a consistency criterion for the model.
            
\begin{figure}
                \centering
                \includegraphics[width = 1\linewidth]{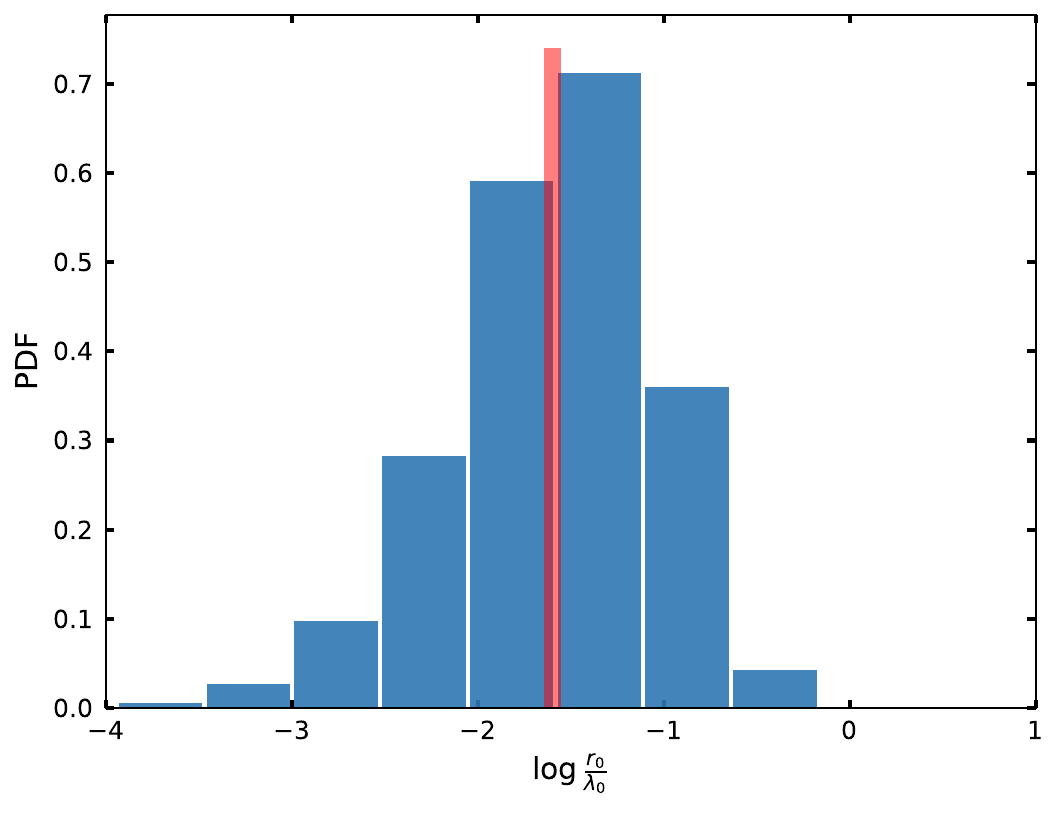}
                \caption{Distribution of the logarithmic mean of the parameter $r_0/\lambda_0$, derived from the distribution of $K_\text{log}$ for sources fitted with the logarithmic helix model. The logarithmic mean of the distribution is indicated by the vertical red line.}
                \label{fig:lambda_ratio_dist}
            \end{figure}

            We conducted Monte Carlo simulations using our estimates of $K_\text{log}$ and the distribution of $\psi$ obtained from the MOJAVE sample \citep{Pushkarev2017}, Fig.~\ref{fig:tgpsi_dist} as a prior. The results are presented in Fig.~\ref{fig:lambda_ratio_dist}. It is seen that the criterion (\ref{eq:lambda_ratio_crit}) is satisfied for the vast majority of the sources. Therefore, we conclude that the resonant mode of the Kelvin--Helmholtz instability model can be used to interpret jet curvatures in AGN. Additionally, we note that given the complications of logarithmic helices modelling for $\alpha > 0$, the actual distribution might be shifted towards even lower values of $\lambda_0 / r_0$, as traced by Fig.~\ref{fig:K_fits}(e), which would further strengthen the KH instability hypothesis.
   
\section{Summary}\label{sec:conclusion}

    We performed a systematic search for curved AGN jets using public VLBI images collected by the Astrogeo database. Starting from 123\,456 intensity images of 17\,478 compact radio sources, we applied an automated filtering procedure to select images with extended structure. The data cover frequencies from 1.2 to 86 GHz, with the largest number of images available at 8 GHz. The filtering reduced the initial data set by 54.6\%, yielding 63\,879 images of 7\,927 sources with resolved jets. Using ridge lines and a simple linear approximation of the position angle variation with distance from the core, we selected candidates with significant curvature and then performed manual validation. As a result, we compiled a final sample of 586 sources classified as having robustly curved pc-scale jets, with an additional 768 sources showing curvature signatures but not included in the robust sample due to insufficient data quality or limited number of observations.
    
    We therefore demonstrate that at least 7.4\% of extended sources show strong apparent curvature on parsec scales, indicating that such curvature is common among resolved VLBI jets. The sample is VLBI-selected, meaning that the observed jets are typically viewed at small angles to the line of sight. Since the curvature is measured in the plane of the sky, projection effects can significantly enhance its apparent magnitude, so the intrinsic curvature may be smaller. Additionally, the bending rate increases with frequency, indicating that the strongest changes in jet direction occur in the innermost regions.

    We find that curved jets are less prevalent at lower redshifts ($z<1$) and more prevalent at higher redshifts ($z>1$). This might arise from selection effects due to poorer linear resolution at larger distances and stronger projection effects in more strongly beamed, small-viewing-angle sources.
    
    Assuming that the observed curvature is produced by jet precession and ballistic motion, the inferred periods range from decades to thousands of years, with a logarithmic mean of $218.8$~years. Disc-driven precession provides a physically consistent explanation for the observed bending. Thin-disc Lense--Thirring precession is compatible with the inferred periods over a broad parameter range, but generally favours relatively high accretion rates, $\dot{m}\sim 0.02\div10$ for $\eta_{0.1}\sim1$, extending from moderately sub-Eddington to super-Eddington regimes. Thick-disc Lense--Thirring precession is less restrictive and naturally produces periods comparable to those inferred here, whereas precession associated with inner-disc warping is viable only within a narrow region near the ADAF boundary. Supermassive black hole binary-driven geodetic precession requires extremely compact, nearly equal-mass binaries, while solid-body disc precession induced by a secondary supermassive black hole remains broadly viable but is constrained by the allowed disc size and binary mass ratio. The resonant Kelvin--Helmholtz instability mode is also consistent with the observations, although no detailed uniform analysis of this model was performed across the sample.
 
    Several key source-specific parameters were not directly constrained, including component kinematics, multi-frequency core shifts, and individual viewing angles. As a result, the present Bayesian analysis cannot decisively distinguish between precession- and instability-driven curvature. Component kinematics would provide a particularly strong diagnostic \citep[e.g. as done in][]{sbj3}. Ballistic components moving along approximately straight trajectories while their ejection direction changes with time would strongly support jet precession. Conversely, components or ridge-line patterns propagating along curved trajectories would disfavour a purely ballistic precession model and support a dynamically evolving structure such as a Kelvin--Helmholtz instability. Measurements of the jet velocity and viewing angle would then help identify the preferred model and constrain its parameters for individual sources.

\begin{acknowledgements}

The authors thank the anonymous referee for their comments and suggestions, which helped improve the quality of the manuscript.
We are deeply grateful for valuable discussions and comments from Galina Lipunova, Yannis Liodakis, Andrei Lobanov, Aleksei Nikonov, Ilya Pashchenko, Alexander Plavin and Elena Shablovinskaya.
This research was funded by the European Union (ERC MuSES project No 101142396). Views and opinions expressed are however those of the author(s) only and do not necessarily reflect those of the European Union or the European Research Council. Neither the European Union nor the granting authority can be held responsible for them.
The work of ABP was supported by the Russian Science Foundation, project 26-12-00190.
A substantial part of this work was carried out during VAM's employment at the Astro Space Center of the Lebedev Physical Institute.
In this work, we used the Astrogeo VLBI FITS image database, maintained by Leonid Petrov.
This database played a crucial role in our work, and we are profoundly grateful to all authors for their significant contributions in providing public VLBI data. 
This research made use of the data from the MOJAVE database maintained by the MOJAVE team \citep{Lister2018}.

This work made use of the Python libraries Matplotlib \citep{Hunter2007}, SciPy \citep{scipy}, Astropy \citep{astropy:2013, astropy:2018, astropy:2022}, and NumPy \citep{numpy}.
\end{acknowledgements}

\section*{Data Availability}

The total intensity images of AGN jets used in our work are publicly available in the Astrogeo VLBI FITS image database 
\href{https://doi.org/10.25966/kyy8-yp57}{DOI: 10.25966/kyy8-yp57}.
The full electronic versions of Table~\ref{tab:curved_sources}, containing the curved-jet classifications, together with the single frequency curvature statistics (Table~\ref{tab:bending_freq}), best-fit parameters for the linear helix model (Table~\ref{tab:jet_hel_fits}) and the logarithmic helix model (Table~\ref{tab:jet_loghel_fits}), are available electronically from CDS.

\bibliographystyle{aa}
\bibliography{curved_jet}

\begin{appendix}
\section{Model-fitting procedure}
\label{sec:model_fitting_appendix}
        
            To estimate physical parameters of the model we first fit helix equations (Eq.~\ref{eq:helix_lin} if $\alpha \sim 1$ or Eq.~\ref{eq:helix_lin_0alpha} $\alpha \ll 1$) to all ridgelines stacked together, aligned by the brightest pixel position. For fits assuming $\alpha \sim 1$, we use a feed-forward modelling approach. Following this approach, we first set an initial guess for the model parameters and then estimate the maximum possible value of the intrinsic rotation angle $\phi$: $\phi_{\text{max}} = \varkappa r_{\text{max}} / K(\alpha + 1)$, where $\varkappa \sim 1.5$ is a parameter that ensures that the model extends beyond the data and $r_\text{max}$ is the maximum distance from the core to the ridge-line points in the image plane. Next, we simulate the guess of the model with $\phi$ between  $0$ and $\phi_{\text{max}}$. To compare the actual data with our guess model, for each ridge-line point we search for the corresponding modelled point. To do so, we search for the point which is the closest to the concentric circle on which our actual data point lies. Next, we compute the reduced $\chi^2$ metric:
            \begin{equation}
                \chi_\text{red}^2 = \frac{1}{N - N_p}\sum_i \left(\frac{|\textbf{r}_i - \textbf{r}_{\text{guess, i}}|}{r_i \cdot \sigma_{\phi, i}}\right)^2,
            \end{equation}
            where $\textbf{r}_i$ -- vector from map centre to the i-th ridgeline point, $\textbf{r}_{\text{guess, i}}$ -- corresponding guess-model points, $\sigma_{\phi, i}$ -- errors of the ridge line's point, $N$ -- total number of stacked ridge lines' points and $N_p$ is a total number of model parameters. To optimise the guess-model we find the global minimum of $\chi^2$. For this purpose we use a non-gradient optimisation routine as implemented in the Python library \texttt{dlib}\footnote{\url{http://dlib.net}}.  
                  
           The optimisation step requires knowledge of the allowed boundaries for each parameter. To define the lower boundary for the parameter \(K\), we use the inequality \(r_\text{max} \leq \sqrt{2} K \phi_\text{max} (1 + \alpha)\), which is obtained from Eq.~\ref{eq:helix_lin}. The order of magnitude of the maximum ridge line length observed in our data across all of the available frequencies is \(r_{ \max}\sim 10\div20\)~mas, and we assume the observed \(\phi_{\max}\) to be between 6$^\circ$ and 180$^\circ$. Hence, the lower boundary on \(K\) is \(K \sim 10^{-3}\)~mas. The upper boundary on \(K\) is set based on the assumption that we still observe the helix with a very small rotation angle \(\phi_{\max}\). The lower boundary on \(\alpha\) is set to be 0.1, because at approximately this value we can no longer distinguish the helix from the case when \(\alpha = 0\). The upper boundary is obtained from the MOJAVE data on opening angles and observation angles \citep{Pushkarev2017}, assuming that they do represent the width and viewing angle of the observed helix.

           Similarly, we obtain parameter constraints for the logarithmic model. In that case, we have an analogous inequality for $r_\text{max}$: $r_\text{max} \leq \sqrt{2}\exp{\left(\phi_\text{max} / K\right)} (1 + \alpha)$, which leads to a maximum \(K\) limit of $10$ and a minimum of $10^{-6}$. The limits for $\alpha$ remain unchanged. 
           
           We fit the data both for models with $\alpha = 0$ and $\alpha \gg 0$. To estimate the uncertainties, we evaluate the inverse Hessian $H^{-1}$ of $\chi^2_\text{red}$ at the optimum, which allows us to directly estimate the parameter covariance matrix $\text{Cov} = 2 H^{-1}/(N - N_p)$.

\section{Generalised Lambda Distribution Fitting} \label{sec:gld}

\begin{figure}
                \centering
                \includegraphics[width = 1\linewidth]{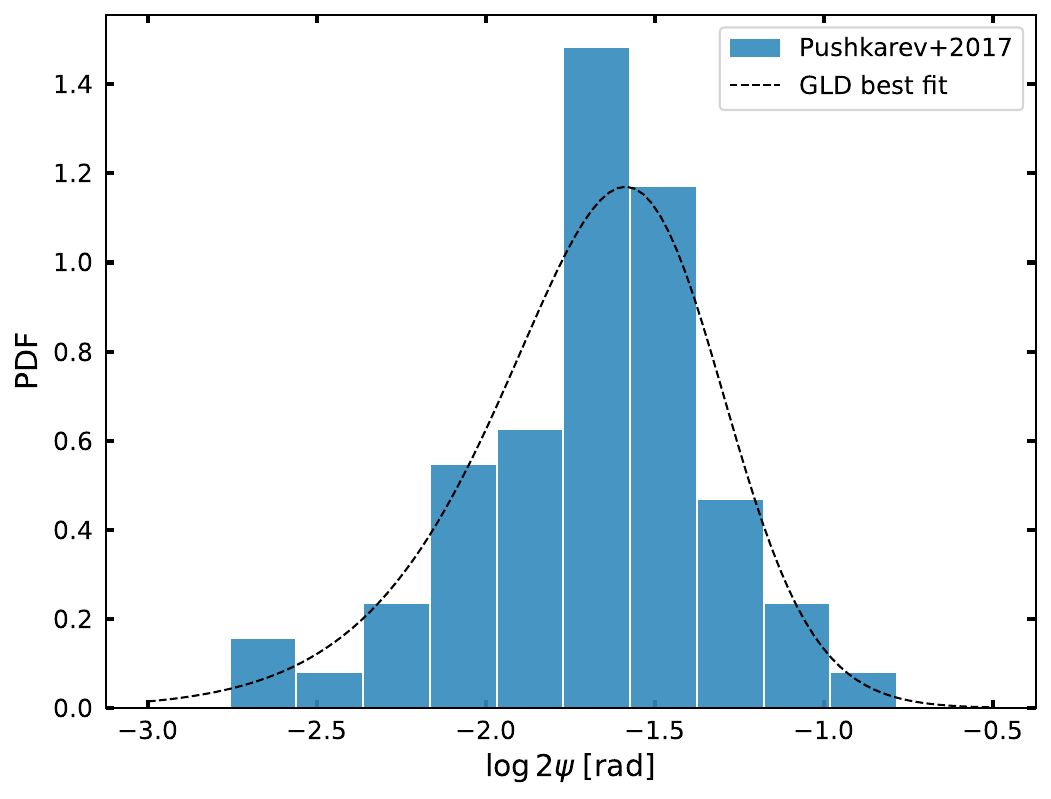}
                \caption{Best fit of the GLD to the estimated distribution of the intrinsic opening angle $2\psi$ in the MOJAVE sample \citep{Pushkarev2017}. The histogram shows the empirical data, and the dashed line represents the fitted GLD. }
                \label{fig:tgpsi_dist}
            \end{figure}

\begin{figure}
                \centering
                \includegraphics[width = 1\linewidth]{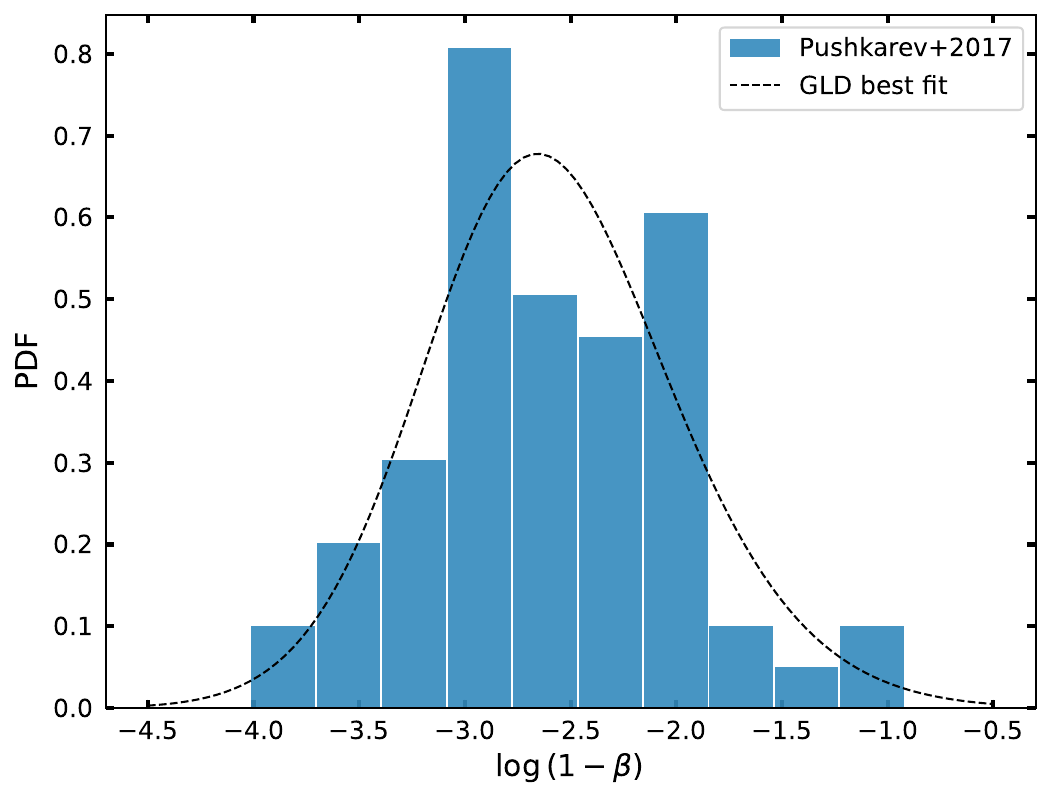}
                \caption{Best fit of the GLD to the estimated distribution of the unit-less intrinsic velocity $\beta = v/c$ in the MOJAVE sample \citep{Pushkarev2017}, where $v$ is the intrinsic velocity and $c$ is the speed of light. The histogram shows the empirical data, and the dashed line represents the fitted GLD. }

                \label{fig:beta_dist}
            \end{figure}

In this appendix, we present the results of fitting several observed distributions using the Generalised Lambda Distribution (GLD) \citep{Karian2000}. The fits were performed using the method of moments as implemented in the Python package \texttt{gldpy}\footnote{\href{https://pypi.org/project/gldpy/}{https://pypi.org/project/gldpy/}}. The goodness-of-fit for each distribution is evaluated using the Kolmogorov--Smirnov (KS) test. For the $\psi$-angle distribution, the KS statistic is 0.0967 with $p=0.544$, while for the $\beta$ distribution it is 0.0689 with $p=0.901$. Fig.~\ref{fig:tgpsi_dist} shows the best fit to the observed distribution of the half-opening angle $\psi$, while Fig.~\ref{fig:beta_dist} shows the distribution of $1-\beta$, where $\beta$ is the intrinsic jet velocity in units of the speed of light. Both datasets are taken from \citet{Pushkarev2017}.

In all cases, the KS test yields high $p$-values, indicating that the GLD provides a good approximation of the empirical distributions. These fits are used as reference distributions for the sampling procedures described in \autoref{subsec:MCmodelling}.

\end{appendix}

\label{lastpage}
\end{document}